\documentclass[longauth]{aa} 

\usepackage{url}
\usepackage{graphicx}
\usepackage{txfonts} 
\usepackage{hyperref} 
\hypersetup{breaklinks=true,colorlinks=true,linkcolor=blue,citecolor=blue,urlcolor=blue}

\usepackage{natbib,twoopt}
\bibpunct{(}{)}{;}{a}{}{,}             

\makeatletter
  \newcommandtwoopt{\citeads}[3][][]{\href{http://adsabs.harvard.edu/abs/#3}%
    {\def\hyper@linkstart##1##2{}%
     \let\hyper@linkend\@empty\citealp[#1][#2]{#3}}}
  \newcommandtwoopt{\citepads}[3][][]{\href{http://adsabs.harvard.edu/abs/#3}%
    {\def\hyper@linkstart##1##2{}%
     \let\hyper@linkend\@empty\citep[#1][#2]{#3}}}
  \newcommandtwoopt{\citetads}[3][][]{\href{http://adsabs.harvard.edu/abs/#3}%
    {\def\hyper@linkstart##1##2{}%
     \let\hyper@linkend\@empty\citet[#1][#2]{#3}}}
  \newcommandtwoopt{\citeyearads}[3][][]%
    {\href{http://adsabs.harvard.edu/abs/#3}
    {\def\hyper@linkstart##1##2{}%
     \let\hyper@linkend\@empty\citeyear[#1][#2]{#3}}}
\makeatother

\graphicspath{{figures/}}

\defcitealias{Husemann:2013}{H13}
\defcitealias{Sanchez:2012a}{S12}
\defcitealias{Walcher:2014}{W14}

\begin{document} 

\titlerunning{The CALIFA survey III. Second public data release}
\authorrunning{Garc\'ia-Benito et al.}
\title{CALIFA, the Calar Alto Legacy Integral Field Area survey}
\subtitle{III. Second public data release\thanks{Based on observations collected 
at the Centro Astron\'omico Hispano Alem\'an (CAHA) at Calar Alto, operated jointly 
by the Max-Planck-Institut f\"ur Astronomie (MPIA) and the Instituto de Astrof\'isica de
Andaluc\'ia (CSIC)}}

\author{R.~Garc\'ia-Benito\inst{\ref{inst1}}
   \and S.~Zibetti\inst{\ref{inst2}}
   \and S.~F.~S\'anchez\inst{\ref{inst3}}
   \and B.~Husemann\inst{\ref{inst4}}
   %
   \and A.~L.~de~Amorim\inst{\ref{inst15}}
   \and A.~Castillo-Morales\inst{\ref{inst29}}
   \and R.~Cid~Fernandes\inst{\ref{inst15}}
   \and S.~C~.~Ellis\inst{\ref{inst21}} 
   \and J.~Falc\'on-Barroso\inst{\ref{inst12},\ref{inst13}}
   \and L.~Galbany\inst{\ref{inst5},\ref{inst6}}
   \and A.~Gil~de~Paz\inst{\ref{inst29}} 
   \and R.~M.~Gonz\'alez Delgado \inst{\ref{inst1}}
   \and E.~A.~D.~Lacerda\inst{\ref{inst15}}
   \and R.~L\'opez-Fernandez\inst{\ref{inst1}}
   \and A.~de~Lorenzo-C{\'a}ceres\inst{\ref{inst11}}
   \and M.~Lyubenova\inst{\ref{inst35},\ref{inst7}}
   \and R.~A.~Marino\inst{\ref{inst40}}
   \and D.~Mast\inst{\ref{inst33}}
   \and M.~A.~Mendoza\inst{\ref{inst1}}
   \and E.~P\'erez\inst{\ref{inst1}}
   \and N.~Vale~Asari\inst{\ref{inst15}}
   %
   %
   \and J.~A.~L.~Aguerri\inst{\ref{inst12},\ref{inst13}}
   \and Y.~Ascasibar\inst{\ref{inst32}}
   \and S.~Bekerait{\.e}\inst{\ref{inst9}}
   \and J.~Bland-Hawthorn\inst{\ref{inst18}}
   \and J.~K.~Barrera-Ballesteros\inst{\ref{inst12},\ref{inst13}}
   \and M.~Cano-D\'iaz\inst{\ref{inst3}}
   \and C.~Catal{\'a}n-Torrecilla\inst{\ref{inst29}} 
   \and C.~Cortijo\inst{\ref{inst1}} 
   \and G.~Delgado-Inglada\inst{\ref{inst3}}
   \and M.~Demleitner\inst{\ref{inst14}} 
   \and R.-J.~Dettmar\inst{\ref{inst30},\ref{inst31}}
   \and A.~I.~D\'iaz\inst{\ref{inst32}}
   \and E.~Florido\inst{\ref{inst22},\ref{inst23}}
   \and A.~Gallazzi\inst{\ref{inst2},\ref{inst26}}
   \and B.~Garc\'ia-Lorenzo\inst{\ref{inst12},\ref{inst13}}
   \and J.~M.~Gomes\inst{\ref{inst17}}
   \and L.~Holmes\inst{\ref{inst27}}  
   \and J.~Iglesias-P\'aramo\inst{\ref{inst1},\ref{inst8}}
   \and K.~Jahnke\inst{\ref{inst7}}
   \and V.~Kalinova\inst{\ref{inst36}}
   \and C.~Kehrig\inst{\ref{inst1}}
   \and R.~C.~Kennicutt Jr\inst{\ref{inst24}}
   \and \'A.~R.~L\'opez-S\'anchez\inst{\ref{inst19},\ref{inst20}}
   \and I.~M\'arquez\inst{\ref{inst1}}
   \and J.~Masegosa\inst{\ref{inst1}}
   \and S.~E.~Meidt\inst{\ref{inst7}}
   \and J.~Mendez-Abreu\inst{\ref{inst11}}
   \and M.~Moll{\'a}\inst{\ref{inst39}} 
   \and A.~Monreal-Ibero\inst{\ref{inst16}}
   \and C.~Morisset\inst{\ref{inst3}}
   \and A.~del~Olmo\inst{\ref{inst1}}
   \and P.~Papaderos\inst{\ref{inst17}}
   \and I.~P\'erez\inst{\ref{inst22},\ref{inst23}}
   \and A.~Quirrenbach\inst{\ref{inst34}}
   \and F.~F.~Rosales-Ortega\inst{\ref{inst28}}
   \and M.~M.~Roth\inst{\ref{inst9}}
   \and T.~Ruiz-Lara\inst{\ref{inst22},\ref{inst23}}
   \and P.~S\'anchez-Bl\'azquez\inst{\ref{inst32}}
   \and L.~S\'anchez-Menguiano\inst{\ref{inst1},\ref{inst22}}
   \and R.~Singh\inst{\ref{inst7}} 
   \and K.~Spekkens\inst{\ref{inst27}}
   \and V.~Stanishev\inst{\ref{inst37},\ref{inst38}}
   \and J.~P.~Torres-Papaqui\inst{\ref{inst25}}
   \and G.~van~de~Ven\inst{\ref{inst7}}
   \and J.~M.~Vilchez\inst{\ref{inst1}}
   \and C.~J.~Walcher\inst{\ref{inst9}}
   \and V.~Wild\inst{\ref{inst11}}
   \and L.~Wisotzki\inst{\ref{inst9}}
   \and B.~Ziegler\inst{\ref{inst10}}
   %
   %
   \and J.~Aceituno\inst{\ref{inst8}}
}

\institute{Instituto de Astrof\'isica de Andaluc\'ia (IAA/CSIC), Glorieta de la Astronom\'{\i}a s/n Aptdo. 3004, E-18080 Granada, Spain,\label{inst1}
\email{rgb@iaa.es}
  \and INAF-Osservatorio Astrofisico di Arcetri - Largo Enrico Fermi, 5 - I-50125 Firenze, Italy\label{inst2}
  \and Instituto de Astronom\'ia, Universidad Nacional Auton\'oma de M\'exico, A.P. 70-264, 04510, México, D.F.\label{inst3}
  \and European Southern Observatory, Karl-Schwarzschild-Str. 2, D-85748 Garching b. M{\"u}nchen, Germany\label{inst4}
  \and Departamento de F\'{\i}sica, Universidade Federal de Santa Catarina, P.O. Box 476, 88040-900, Florian\'opolis, SC, Brazil\label{inst15} 
  \and Departamento de Astrof{\'i}sica y CC. de la Atm{\'o}sfera, Universidad Complutense de Madrid, E-28040, Madrid, Spain\label{inst29}
  \and Australian Astronomical Observatory, 105 Delhi Road, North Ryde, NSW 2113, Australia\label{inst21}
  \and Instituto de Astrof\'isica de Canarias, V\'ia L\'actea s/n, La Laguna, Tenerife, Spain\label{inst12}
  \and Departamento de Astrof\'isica, Universidad de La Laguna, E-38205 La Laguna, Tenerife, Spain\label{inst13}
  \and Millennium Institute of Astrophysics, Universidad de Chile, Santiago, Chile\label{inst5}
  \and Departamento de Astronom\'ia, Universidad de Chile, Casilla 36-D, Santiago, Chile\label{inst6}
  \and School of Physics and Astronomy, University of St Andrews, SUPA, North Haugh, KY16 9SS, St Andrews, UK \label{inst11}
  \and Kapteyn Astronomical Institute, University of Groningen, Postbus 800, 9700 AV Groningen, The Netherlands\label{inst35}
  \and Max-Planck-Institut f\"ur Astronomie, K\"onigstuhl 17, D-69117 Heidelberg, Germany\label{inst7}
  \and CEI Campus Moncloa, UCM-UPM, Departamento de Astrof\'{i}sica y CC$.$ de la Atm\'{o}sfera, Facultad de CC$.$ F\'{i}sicas, Universidad Complutense de Madrid, Avda.\,Complutense s/n, 28040 Madrid, Spain\label{inst40}
  \and Instituto de Cosmologia, Relatividade e Astrof\'{i}sica – ICRA, Centro Brasileiro de Pesquisas F\'{i}sicas, Rua Dr.Xavier Sigaud 150, CEP 22290-180, Rio de Janeiro, RJ, Brazil\label{inst33}
  \and Departamento de F\'isica Te\'orica, Facultad de Ciencias, Universidad Aut\'onoma de Madrid, E-28049 Madrid, Spain\label{inst32} 
  \and Leibniz-Institut f\"ur Astrophysik Potsdam (AIP), An der Sternwarte 16, D-14482 Potsdam, Germany\label{inst9}
  \and Sydney Institute for Astronomy, School of Physics, University of Sydney, NSW 2006, Australia\label{inst18}
  \and Universit\"at Heidelberg, Zentrum f\"ur Astronomie, Astronomisches Rechen-Institut, M\"onchhofstra{\ss}e 12-14, D-69120 Heidelberg, Germany\label{inst14}
  \and Astronomisches Institut, Ruhr-Universit{\"a}t Bochum, Universit{\"a}tsstr. 150, D-44801 Bochum, Germany\label{inst30}
  \and RUB Research Department Plasmas with Complex Interactions\label{inst31}
  \and Departamento de F\'{\i}sica Te\'orica y del Cosmos, University of Granada, Facultad de Ciencias (Edificio Mecenas), E-18071 Granada, Spain\label{inst22}
  \and Dark Cosmology Center, University of Copenhagen, Niels Bohr Institute, Juliane Maries Vej 30, 2100 Copenhagen, Denmark\label{inst26}
  \and Instituto de Astrof{\'i}sica e Ci\^{e}ncias do Espa\c{c}o, Universidade do Porto, CAUP, Rua das Estrelas, PT4150-762 Porto, Portugal\label{inst17}
  \and Department of Physics, Royal Military College of Canada, PO box 17000, Station Forces, Kingston, Ontario, Canada, K7K 7B4\label{inst27}
  \and Centro Astron\'omico Hispano Alem\'an de Calar Alto (CSIC-MPG), C/ Jes\'us Durb\'an Rem\'on 2-2, E-4004 Almer\'ia, Spain\label{inst8}
  \newpage
  \and Department of Physics 4-181 CCIS, University of Alberta, Edmonton AB T6G 2E1, Canada\label{inst36}
  \and Institute of Astronomy, University of Cambridge, Madingley Road, Cambridge CB3 0HA, UK\label{inst24}
  \and Australian Astronomical Observatory, PO Box 915, North Ryde, NSW 1670, Australia\label{inst19}
  \and Department of Physics and Astronomy, Macquarie University, NSW 2109, Australia\label{inst20}
  \and CIEMAT, Avda. Complutense 40, 28040 Madrid, Spain\label{inst39}
  \and GEPI, Observatoire de Paris, CNRS, Université Paris Diderot, Place Jules Janssen, 92190 Meudon, France\label{inst16}
  \and Instituto Universitario Carlos I de F\'isica Te\'orica y Computacional, Universidad de Granada, 18071 Granada, Spain\label{inst23}
  \and Landessternwarte, Zentrum f\"ur Astronomie der Universit\"at Heidelberg, K\"onigstuhl 12, D-69117 Heidelberg, Germany\label{inst34}
  \and Instituto Nacional de Astrof{\'i}sica, {\'O}ptica y Electr{\'o}nica, Luis E. Erro 1, 72840 Tonantzintla, Puebla, Mexico\label{inst28}
  \and CENTRA - Centro Multidisciplinar de Astrof\'isica, Instituto Superior T\'ecnico, Av. Rovisco Pais 1, 1049-001 Lisbon, Portugal\label{inst37}
  \and Department of Physics, Chemistry and Biology, IFM, Link\"oping University, SE-581 83 Link\"oping, Sweden\label{inst38}
  \and Departamento de Astronom\'ia, Universidad de Guanajuato, Apartado Postal 144, 36000, Guanajuato, Guanajuato, Mexico\label{inst25}
  \and University of Vienna, Department of Astrophysics, T\"urkenschanzstr. 17, 1180 Vienna, Austria\label{inst10}
}

\date{}

 
  \abstract
   {This paper describes the Second Public Data Release (DR2) of the Calar Alto Legacy Integral Field Area (CALIFA) 
   survey. The data for 200 objects are made public, including the 100 galaxies of the First Public Data Release (DR1). 
   Data were obtained with the integral-field spectrograph PMAS/PPak mounted on the 3.5 m telescope at the Calar Alto 
   observatory. 
   Two different spectral setups are available for each galaxy, (i) a low-resolution V500 setup covering the wavelength 
   range 3745–7500 \AA\ with a spectral resolution of 6.0 \AA\ (FWHM), and (ii) a medium-resolution V1200 setup covering 
   the wavelength range 3650–4840 \AA\ with a spectral resolution of 2.3 \AA\ (FWHM). The sample covers a redshift range 
   between 0.005 and 0.03, with a wide range of properties in the Color-Magnitude diagram, stellar mass, 
   ionization conditions, and morphological types. All released cubes were reduced with the latest pipeline, 
   including improved spectrophotometric calibration, spatial registration and spatial resolution. The spectrophotometric 
   calibration is better than 6\% and the median spatial resolution is 2\farcs5. Altogether the second data release 
   contains over 1.5 million spectra. It is available at \url{http://califa.caha.es/DR2}.} 


\maketitle

\section{Introduction}

The Calar Alto Legacy Integral Field Area (CALIFA) survey \citep[][hereafter S12]{Sanchez:2012a} is an 
ongoing large project of the Centro Astron\'omico Hispano-Alem\'an at the Calar Alto observatory 
(Almer\'ia, Spain) to obtain spatially resolved spectra for 600 galaxies in the Local Universe by means 
of integral field spectroscopy (IFS). CALIFA observations started in June 2010 with the Potsdam Multi 
Aperture Spectrograph \citep[PMAS,][]{Roth:2005}, mounted on the 3.5 m telescope, utilizing the large 
hexagonal field-of-view (FoV) offered by the PPak fiber bundle 
\citep{Verheijen:2004,Kelz:2006}. Each galaxy is observed using two different setups: an intermediate 
spectral resolution one (V1200, $R\sim 1650$) and a low-resolution one (V500, $R\sim 850$).
A diameter-selected sample of 939 galaxies was drawn from the 7th data release of the Sloan Digital 
Sky Survey \citep[SDSS DR7,][]{Abazajian:2009} which is described in \citet[hereafter W14]{Walcher:2014}. 
From this mother sample the 600 target galaxies are randomly selected.

Combining the techniques of imaging and spectroscopy through optical IFS provides a more comprehensive 
view of individual galaxy properties than any traditional survey. CALIFA-like observations were collected
during the feasibility studies \citep{MarmolQueralto:2011,Viironen:2012} and the PPak IFS Nearby Galaxy 
Survey \citep[PINGS,][]{RosalesOrtega:2010}, a predecessor of this survey. First results based on 
those datasets already explored their information content \citep[e.g.,][]{Sanchez:2011,
RosalesOrtega:2011,AlonsoHerrero:2012,Sanchez:2012b,RosalesOrtega:2012}. CALIFA can therefore 
be expected to make a substantial contribution to our understanding of galaxy evolution in various
aspects, including (i) the relative importance and consequences of merging and secular processes; 
(ii) the evolution of galaxies across the color-magnitude diagram; (iii) the effects of the environment 
on galaxies; (iv) the AGN--host galaxy connection; (v) the internal dynamical processes in galaxies; 
and (vi) the global and spatially resolved star formation history and chemical enrichment of various 
galaxy types.

Compared with previous IFS surveys, e.g., Atlas3D \citep{Cappellari:2011} or DMS \citep{Bershady:2010}, 
CALIFA covers a much wider range of morphological types over a large range of masses, sampling the 
entire Color-Magnitude diagram for $M_r > -19$ mag. While the recently started SAMI 
\citep{Croom:2012,Bryant:2014} and MaNGA \citep{Law:2014} surveys have a similarly broad scope 
as CALIFA and aim at building much larger samples, CALIFA has still an advantage in terms of spatial 
coverage and sampling. For 50\% of the galaxies, CALIFA provides data out to 3.5 $r_\mathrm{e}$, 
and for 80\% out to 2.5 $r_\mathrm{e}$. At the same time the spatial resolution of
$\sim 1$ kpc is typically better than in either SAMI or MaNGA, revealing several
of the most relevant structures in galaxies (spiral arms, bars, bulges, giant
\ion{H}{2} regions, etc.). CALIFA has lower spectral resolution than these two
surveys in the red, but is comparable for the blue wavelength range.

So far, a number of science goals have been addressed using the data from the CALIFA survey: (i) New 
techniques have been developed to understand the spatially resolved star formation histories (SFH) of
galaxies \citep{CidFernandes:2013,CidFernandes:2014}. We found solid evidence that mass-assembly 
in the typical galaxies happens from the inside-out \citep{Perez:2013}. The SFH and metal 
enrichment of bulges and early-type galaxies are fundamentally related to the total 
stellar mass, while for disk galaxies it is more 
related to the local stellar mass density \citep{GonzalezDelgado:2014a,GonzalezDelgado:2014b}; 
(ii) We developed new tools to detect and extract the spectroscopic information of \ion{H}{ii} 
\citep{Sanchez:2012b}, building the largest catalog currently available ($\sim$6,000 \ion{H}{ii} 
regions and aggregations). This catalog has been used to define a new oxygen abundance calibrator 
anchored to electron temperature measurements \citep{Marino:2013}. From these, we explored the 
dependence of the mass-metallicity relation with Star Formation Rate \citep{Sanchez:2013}, 
and the local mass-metallicity relation \citep{RosalesOrtega:2012}. We found that all galaxies in our 
sample present a common gas-phase oxygen abundance radial gradient with a similar slope when normalized 
to the effective radius \citep{Sanchez:2014}, which agrees with an inside-out scenario 
for galaxy growth. This characteristic slope is independent of the properties of the galaxies, and 
in particular of the presence or absence of a bar, contrary to previous results. More recently, 
this result has been confirmed by the analysis of the stellar abundance gradient in the
same sample \citep{SanchezBlazquez:2014}; (iii) We explored the origin of the low intensity, 
LINER-like, ionized gas in galaxies. These regions are clearly not related to
star-formation activity, or to AGN activity. They are most probably related to post-AGB 
ionization in many cases \citep{Kehrig:2012,Singh:2013,Papaderos:2013}; (iv) We explored the
aperture and resolution effects on the data. CALIFA provides a unique tool to understand 
the aperture and resolution effects in larger single-fiber (e.g. SDSS) and IFS surveys 
(e.g. MaNGA, SAMI). We explored the effects of the dilution of the signal in different gas
and stellar population properties \citep{Mast:2014}, and proposed a new empirical 
aperture correction for the SDSS data \citep{IglesiasParamo:2013}; 
(v) We have analysed the local properties of the ionized gas and stellar population of 
galaxies where supernovae (SNe) have exploded. Core collapse SNe are found closer to 
younger stellar populations, while SNe Ia show no correlation to stellar 
age \citep{Galbany:2014}; 
(vi) CALIFA is the first IFS survey that allows gas and stellar kinematic studies for 
all morphologies with enough spectroscopic resolution to study (a) the kinematics 
of the ionized gas \citep{GarciaLorenzo:2014}, (b) the effects of bars in the kinematics 
of galaxies \citep{BarreraBallesteros:2014a}; (c) the effects of the interaction stage 
on the kinematic signatures (Barrera-Ballesteros et al., submitted), (d) the Bar Pattern
Speeds in late-type galaxies (Aguerri et al., submitted), (e) the measurements of 
the angular momentum of galaxies to previously unexplored ranges of morphology and 
ellipticity (Falc\'on-Barroso et al., in prep.); (vii) We explored the effects of a first 
stage merger on the gas and stellar kinematics, star formation activity and stellar populations 
of the Mice merging galaxies \citep{Wild:2014}.

In this article, we introduce the second data release (DR2) of CALIFA, which grants public 
access to high-quality data for a set of 200 galaxies (400 datacubes). All released cubes have 
been reduced with the latest pipeline, including improved spectrophotometric calibration, spatial 
registration and spatial resolution. This DR supersedes and increases by a factor of two the 
amount of data delivered in DR1 \citep[][hereafter H13]{Husemann:2013}.

DR1 opened CALIFA to the community, and allowed for the exploration of several different 
scientific avenues not addressed by the collaboration 
\citep[e.g.][]{Holwerda:2013,DeGeyter:2014,MartinezGarcia:2014,Davies:2014}. 
The properties of the galaxies in the DR2 sample are summarized in Sect.~\ref{sect:DR2_sample}. 
We describe the processing (Sect.~\ref{sect:data_processing}), structure 
(Sect.~\ref{sect:data_format}), and data (Sect.~\ref{sect:QC}) of the distributed 
CALIFA data as essential information for any scientific analysis. The several interfaces 
to access the CALIFA DR2 data are explained in Sect.~\ref{sect:DR2_access}.

\section{The CALIFA DR2 sample}\label{sect:DR2_sample}

\begin{figure}
 \resizebox{\hsize}{!}{\includegraphics{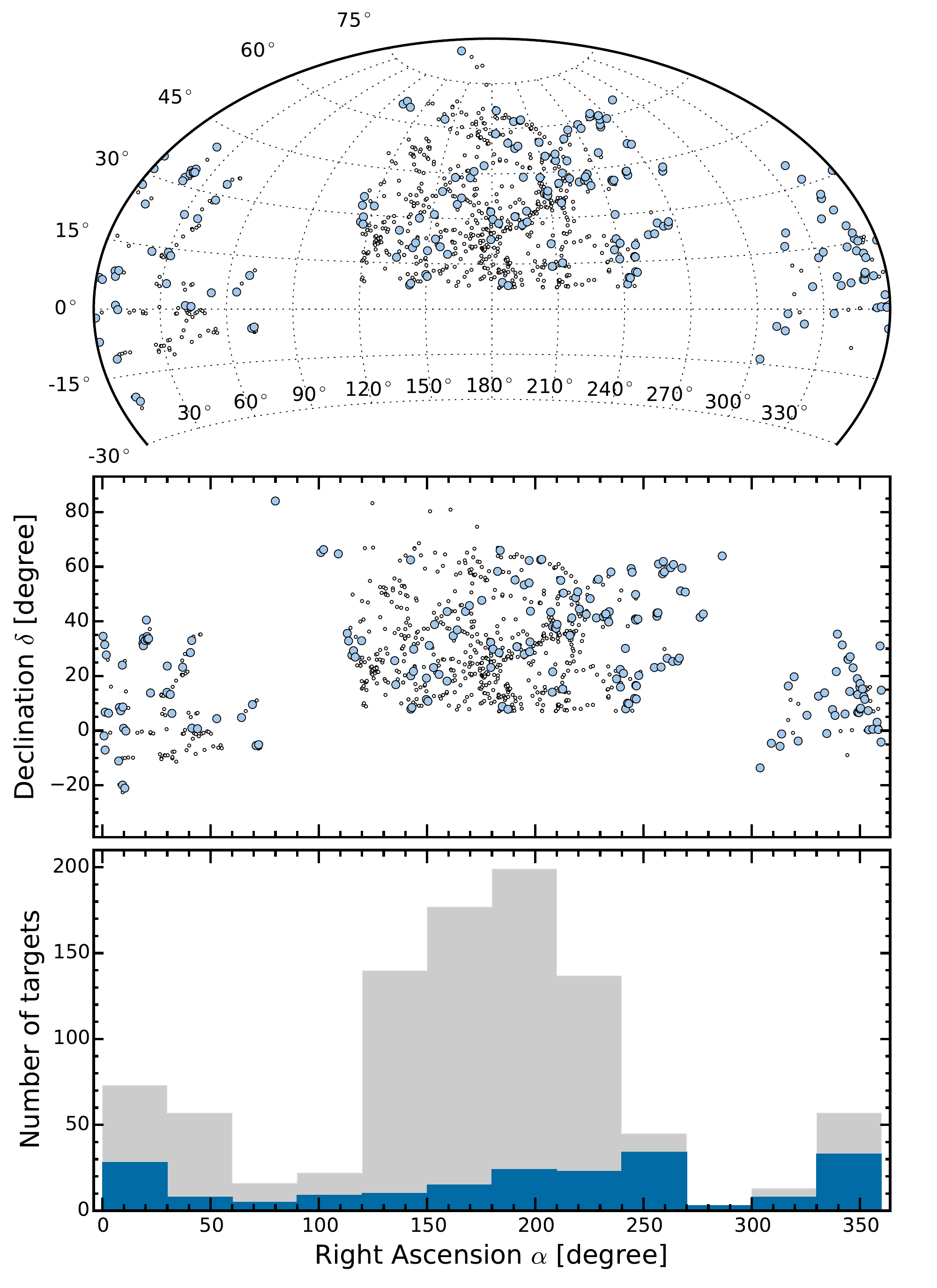}}
  \caption{Distribution on the sky of galaxies in the CALIFA mother sample (black dots) and CALIFA DR2 sample 
   (blue filled symbols). The upper panel shows the distribution in an Aitoff projection in J2000 Equatorial 
   Coordinates (cut off at $\delta$ = $-$30\degr, below which the sample does not extend), while the middle 
   panel is plotted in the cartesian system. The lower panel shows both samples as a function of right 
   ascension. The number distribution in bins of 30\degr\ along the right ascension is shown  
   for the mother sample (grey area) and the DR2 sample (blue area).}
  \label{fig:DR2_dist}
\end{figure}

The CALIFA ``mother sample'' (MS) consists of 939 galaxies drawn from SDSS DR7. The main criteria for 
the target selection are: angular isophotal diameter ($45\arcsec < isoA_{r} < 79.2\arcsec$) of the 
galaxies\footnote{$isoA_{r}$ is the isophote major axis at 25 magnitudes per square arcsecond in the $r$-band. 
For other SDSS pipeline parameters meaning, the reader is referred to the DR7 webpage: 
\url{http://skyserver.sdss.org/dr7/en/help/browser/browser.asp}}; redshift range $0.005<z<0.03$; cut in Galactic 
latitude to exclude the Galactic plane ($|b| > 20\degr$); flux limit of $petroMag_{r} < 20$; 
declination limit to $\delta > 7\degr$. Redshift limits were imposed so that the sample would not be dominated 
by dwarf galaxies and in order to keep relevant spectral features observed with a fixed instrumental spectral 
setup. Redshift information was taken from SIMBAD for all galaxies where SDSS DR7 spectra were unavailable. 
The cut in declination was chosen to reduce problems due to differential atmospheric refraction (DAR) and 
PMAS flexure issues, but was not applied to the SDSS Southern area due to the sparsity of objects in this region. 
The reader is referred to \citetalias{Walcher:2014} for a comprehensive characterization of the CALIFA MS 
and a detailed evaluation of the selection effects implied by the chosen criteria. From the CALIFA MS, 
600 galaxies are randomly selected for observation purely based on visibility, and we refer to these galaxies 
as the virtual final CALIFA sample hereafter. 

\begin{longtab}
\begin{longtable}{lcccccccccc}
\caption{CALIFA DR2 galaxies and their characteristics. }\\\hline\hline
\label{tab:DR2_sample}
\small{Name} &  \small{ID\tablefootmark{a}} & \small{$\alpha$(J2000)\tablefootmark{b}} & \small{$\delta$(J2000)\tablefootmark{b}} & \small{$z$\tablefootmark{c}} & \small{$m_g$\tablefootmark{d}} & \small{$m_z$\tablefootmark{d}} & \small{$m_u-m_z$\tablefootmark{d}} & \small{type\tablefootmark{e}} & \small{bar\tablefootmark{f}}  & \small{$b/a$\tablefootmark{g}}\\\hline
\endfirsthead
\caption{continued.}\\\hline\hline
\small{Name} &  \small{ID\tablefootmark{a}} & \small{$\alpha$(J2000)\tablefootmark{b}} & \small{$\delta$(J2000)\tablefootmark{b}} & \small{$z$\tablefootmark{c}} & \small{$m_g$\tablefootmark{d}} & \small{$m_z$\tablefootmark{d}} & \small{$m_u-m_z$\tablefootmark{d}} & \small{type\tablefootmark{e}} & \small{bar\tablefootmark{f}}  & \small{$b/a$\tablefootmark{g}}\\\hline
\endhead
\hline
\endfoot
\small{\object{IC5376}} & \small{001} & \small{00:01:19.779} & \small{+34:31:32.409} & \small{0.0168} & \small{14.24} & \small{12.60} & \small{ 3.48} & \small{Sb} & \small{A} & \small{0.27} \\
\small{\object{UGC00005}} & \small{002} & \small{00:03:05.643} & \small{-01:54:49.804} & \small{0.0243} & \small{13.88} & \small{12.53} & \small{ 2.95} & \small{Sbc} & \small{A} & \small{0.54} \\
\small{\object{NGC7819}} & \small{003} & \small{00:04:24.505} & \small{+31:28:19.228} & \small{0.0167} & \small{14.06} & \small{13.01} & \small{ 2.12} & \small{Sc} & \small{A} & \small{0.53} \\
\small{\object{IC1528}} & \small{005} & \small{00:05:05.377} & \small{-07:05:36.204} & \small{0.0128} & \small{13.46} & \small{12.52} & \small{ 2.43} & \small{Sbc} & \small{AB} & \small{0.36} \\
\small{\object{UGC00036}} & \small{007} & \small{00:05:13.882} & \small{+06:46:19.306} & \small{0.0210} & \small{14.12} & \small{12.46} & \small{ 3.55} & \small{Sab} & \small{AB} & \small{0.60} \\
\small{\object{NGC0001}} & \small{008} & \small{00:07:15.860} & \small{+27:42:29.096} & \small{0.0151} & \small{13.46} & \small{12.01} & \small{ 2.97} & \small{Sbc} & \small{A} & \small{0.80} \\
\small{\object{NGC0036}} & \small{010} & \small{00:11:22.298} & \small{+06:23:21.667} & \small{0.0203} & \small{13.46} & \small{12.01} & \small{ 3.23} & \small{Sb} & \small{B} & \small{0.65} \\
\small{\object{MCG-02-02-030}} & \small{013} & \small{00:30:07.309} & \small{-11:06:49.066} & \small{0.0118} & \small{13.41} & \small{12.08} & \small{ 2.91} & \small{Sb} & \small{AB} & \small{0.34} \\
\end{longtable}
\tablefoot{
\tablefoottext{a}{CALIFA unique ID number for the galaxy.}
\tablefoottext{b}{Equatorial coordinates of the galaxies as provided by NED.}
\tablefoottext{c}{Redshift of the galaxies based on SDSS DR7 spectra or complemented with SIMBAD information if SDSS spectra are not available.}
\tablefoottext{d}{Petrosian magnitudes as given by SDSS DR7 database corrected for Galactic extinction.}
\tablefoottext{e}{Morphological type from our own visual classification (see \citetalias{Walcher:2014} for details). ``(x)'' indicates ongoing mergers.}
\tablefoottext{f}{Bar strength of the galaxy as an additional outcome of our visual classification. A stands for non-barred, B for barred and AB if unsure.}
\tablefoottext{g}{Ratio between the semi-minor and semi-major axis based on a detailed re-analysis of the SDSS images (see \citetalias{Walcher:2014} for details).}
\tablefoottext{h}{Morphological classification of this particular galaxy NGC 4676B from \cite{Wild:2014}.}
}
\end{longtab}

The 200 DR2 galaxies, which include the first 100 galaxies of DR1, were observed in both spectral setups from 
the start of observations in June 2010 until December 2013. We list these galaxies in Table~\ref{tab:DR2_sample} 
together with their primary characteristics. The distribution of galaxies in the sky follows the underlying 
SDSS footprint (Fig.~\ref{fig:DR2_dist}). The number of galaxies in DR2 is not homogeneous as a function of 
right ascension, $\alpha$(J2000), and has three mean clear peaks at around $\alpha$ $\sim$ 15\degr, 255\degr 
and 345\degr. All three peaks are located in the same season run, in the period from April to October. As noted 
in \citetalias{Husemann:2013}, there was a downtime of the 3.5 m telescope from August 2010 until April 2011 
due to operational reasons at the observatory, which delayed the survey roughly by 8 months. In addition to this, 
due to scheduling matters, a large part of the granted time was allocated in summer seasons. Regardless of the 
observing time issue, the distribution of physical properties for DR2 is nearly random, as expected, 
and covers galaxies with a wide range of properties as discussed below.

\begin{figure}
 \resizebox{\hsize}{!}{\includegraphics{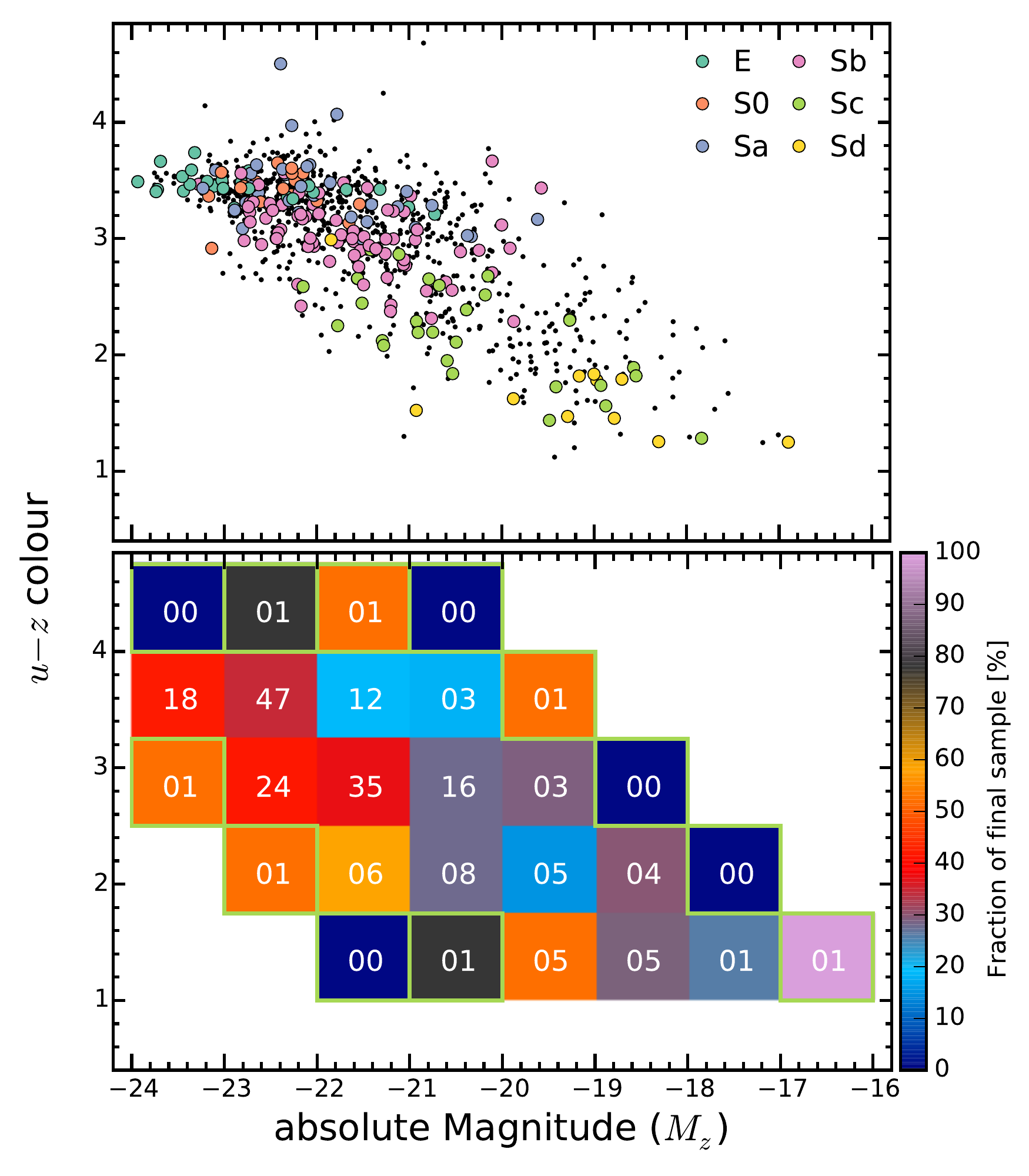}}
  \caption{\textit{Upper panel:} Distribution of CALIFA galaxies in the $u-z$ vs. $M_z$ color-magnitude diagram. 
    Black dots denote galaxies in the CALIFA mother sample (\citetalias{Sanchez:2012a}, \citetalias{Walcher:2014}) 
    and colored symbols indicate CALIFA DR2 galaxies. Different colors account for the morphological classification, 
    which range from ellipticals (E) to late-type spirals (Sd). \textit{Lower panel:} Fraction of 
    galaxies in the DR2 sample with respect to the expected final CALIFA sample distribution (600 objects) in 
    bins of 1\,mag in $M_z$ and 0.75\,mag in $u-z$. The total number of galaxies per bin in the DR2 sample is 
    shown in each bin. Bins for which the number of galaxies in the mother sample is less than 5 are prone to 
    low-number statistics and enclosed by a green square for better identification.}
  \label{fig:DR2_CMD}
\end{figure}

\begin{figure}
\resizebox{\hsize}{!}{\includegraphics{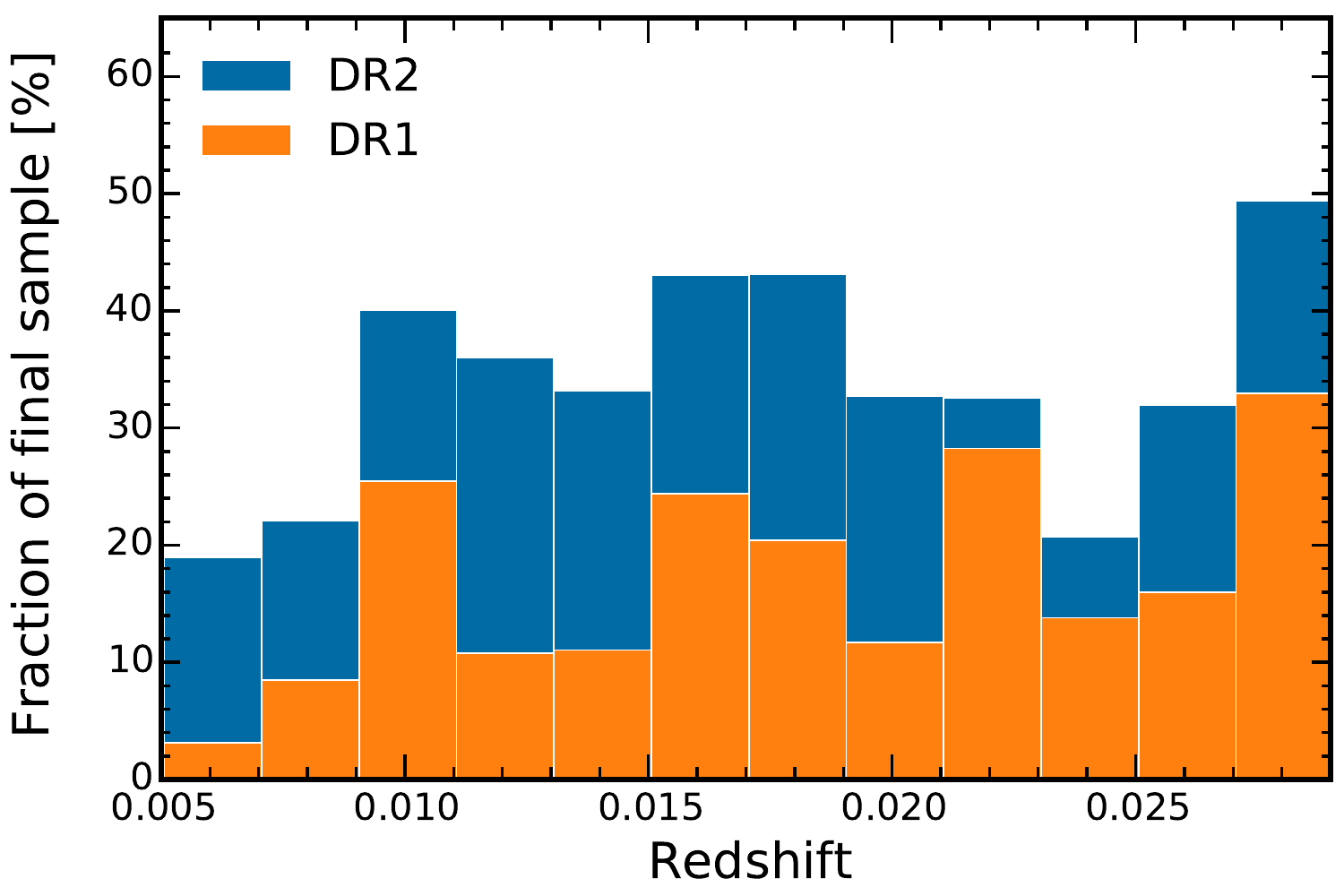}}
 \caption{Redshift distribution of the DR2 (blue) and DR1 (orange) as a fraction of the final CALIFA sample.}
 \label{fig:DR2_redshift}
\end{figure}

\begin{figure}
\resizebox{\hsize}{!}{\includegraphics{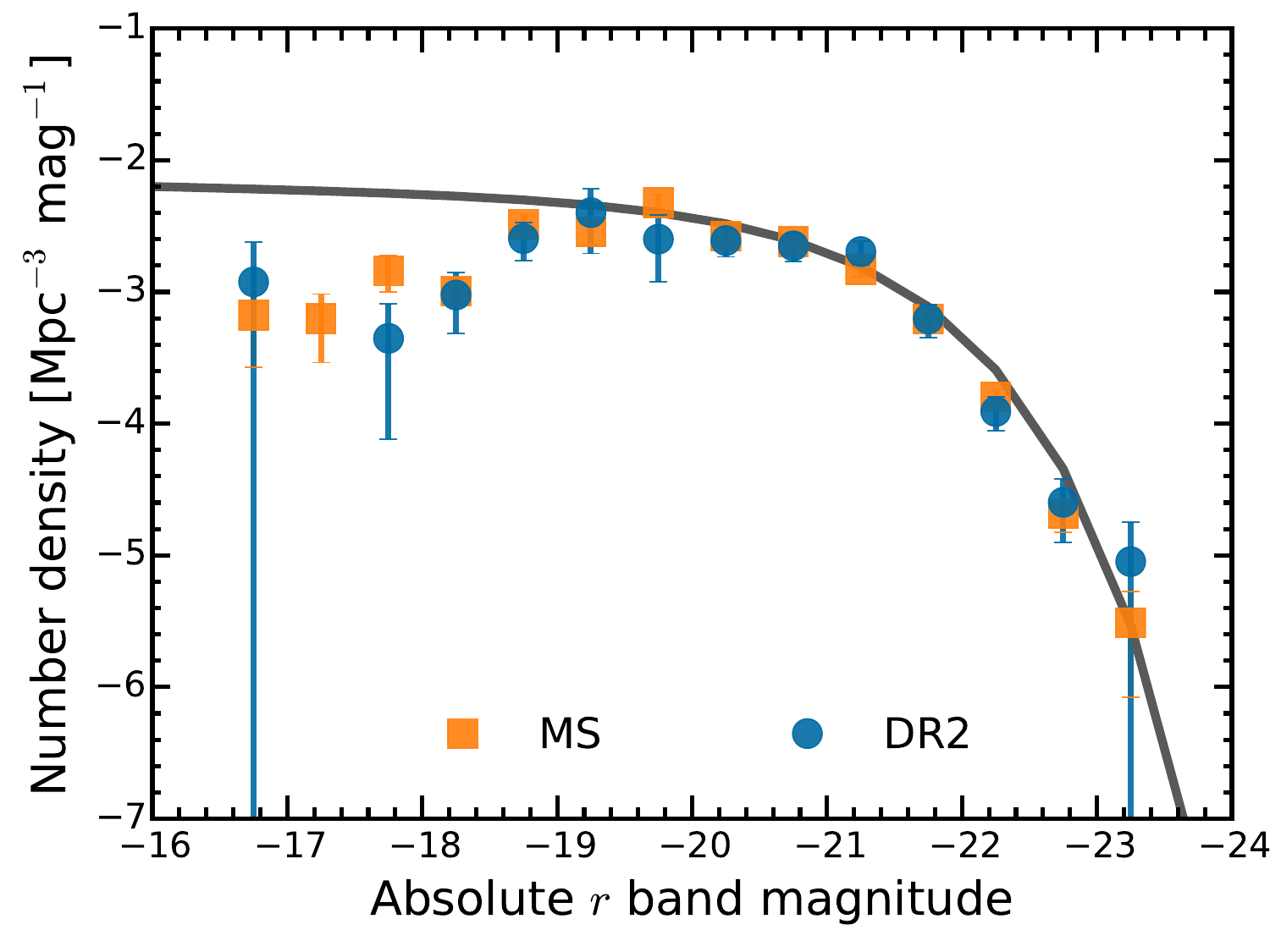}}
 \caption{Luminosity functions in the $r$  band of the CALIFA mother sample (orange squares) and the DR2 sample 
     (blue points). Error bars represent Poissonian uncertainties. The line shows the Schechter fit to the LF of 
     \citet{Blanton:2005}.}
 \label{fig:DR2_LF}
\end{figure}

Figure \ref{fig:DR2_CMD} shows the distribution of galaxies in the color-magnitude diagram. It is evident that 
the DR2 sample covers nearly the full range of the CALIFA MS. On average, the DR2 targets comprise $\sim$37\% per 
color-magnitude bin of the total 600 objects when CALIFA is completed. The deficit of low luminosity galaxies with 
intermediate colors noted in DR1 has improved. Fluctuations can be explained by the effect of low 
number statistics, especially within those color-magnitude bins in which the MS contains fewer galaxies. 
This point is highlighted in Figure~\ref{fig:DR2_CMD} and emphasizes the need to increase further the numbers 
to the full CALIFA sample to obtain enough galaxies in each bin for a meaningful multi-dimensional statistical 
analysis. 

Figure \ref{fig:DR2_redshift} compares the redshift distribution of the CALIFA galaxies in the DR2 and DR1, as 
a fraction of the CALIFA sample. As it can be seen, except for a few particular bins, the redshift 
distribution is homogeneous with respect to the final sample.

One important test to be made is whether the number density of galaxies estimated from the CALIFA sample is in 
accordance with other surveys. Figure~\ref{fig:DR2_LF} shows the $r$-band luminosity function (LF) of the DR2 
sample as compared to the MS and the reference SDSS sample of \citet{Blanton:2005}. The reader is 
referred to \citetalias{Walcher:2014} for all technical details on how the LFs is obtained and for the 
explanation of the turnover of the LF at $M_r \approx -18.6$. It should be noted that the DR2 sample already 
reproduces very closely the CALIFA MS LF in most of its magnitude bins. 

\begin{figure}
 \resizebox{\hsize}{!}{\includegraphics{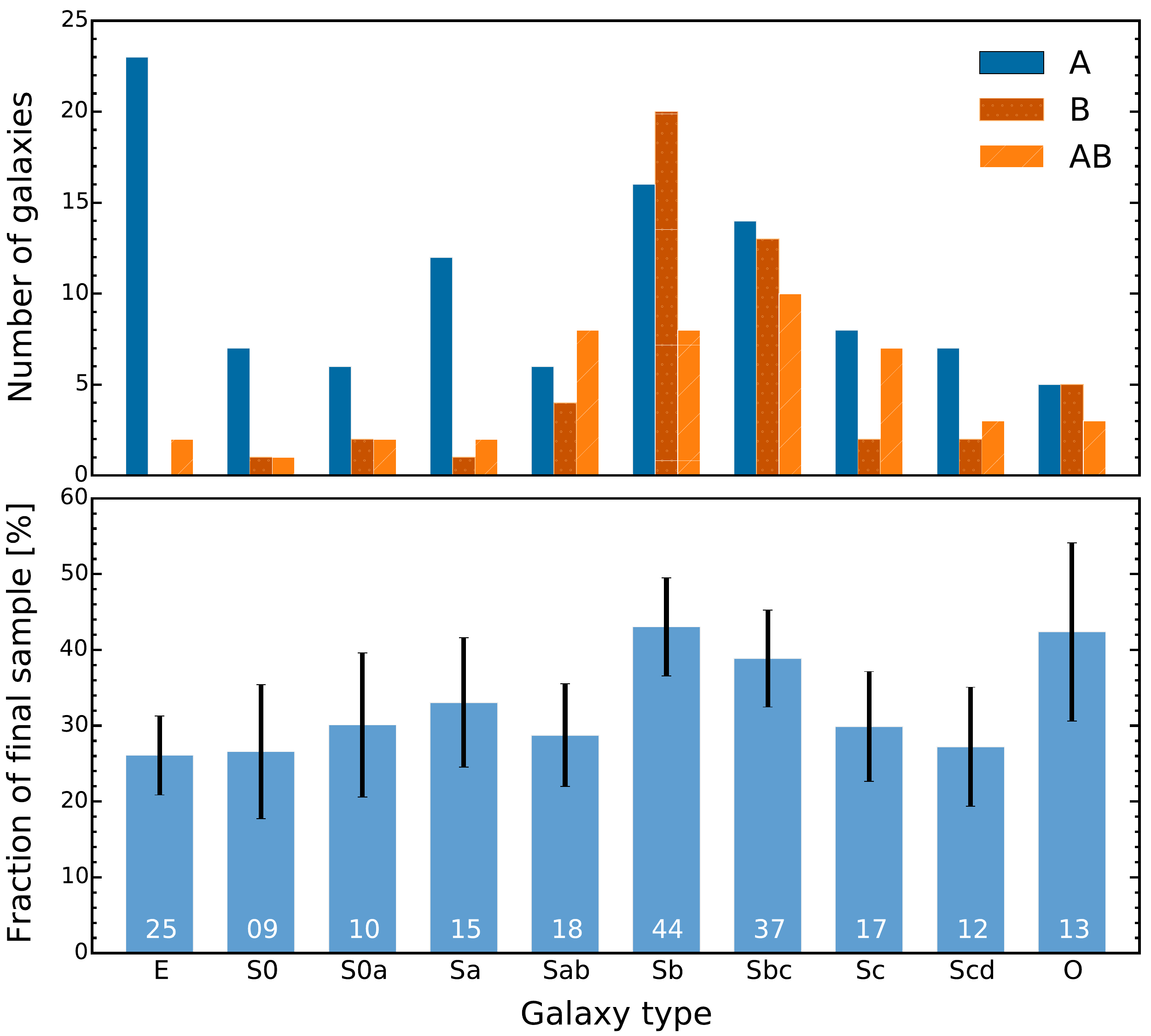}}
  \caption{The distribution of visually classified morphological types in the DR2 sample. We divide the galaxies 
        into ellipticals (E), spirals (from S0 to Scd) and the \textit{other} 
        group ``O'' which includes Sd, Sdm, Sm and I (only one) types. \textit{Upper panel:} Bar strength 
	histogram, where A stands for non-barred, B for barred and AB if unsure.  \textit{Lower panel:} 
        The fraction of galaxies in the DR2 sample with respect to the expected final CALIFA sample distribution.
	The total number of galaxies in the DR2 for each morphology type is written on each bar. Error bars are 
	computed from the Poisson errors of the associated DR2 number counts. The morphological distribution of 
        the DR2 sample is similar to that of the mother sample.}
  \label{fig:DR2_morph}
\end{figure}

\begin{figure}
 \resizebox{\hsize}{!}{\includegraphics{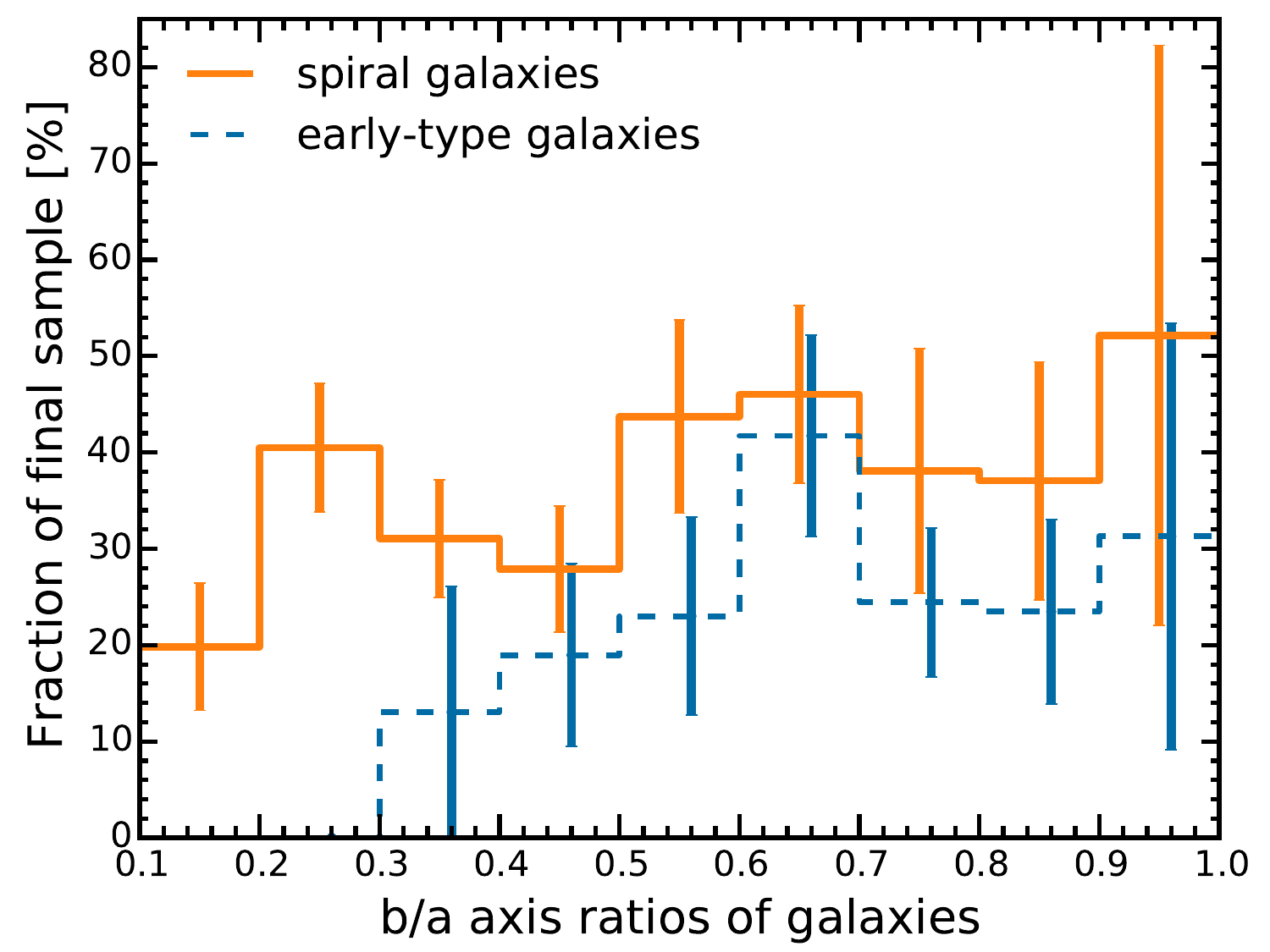}}
  \caption{The fraction of galaxies in the DR2 sample with respect to the expected final CALIFA sample distribution, 
    as a function of the light-weighted axis ratio ($b/a$). Galaxies were separated into early-type galaxies (E+S0) 
    and spiral galaxies (Sa and later). The CALIFA mother sample does not include any elliptical galaxies 
    with $b/a<0.3$. Error bars are computed from the Poisson errors of the associated DR2 number counts.}
  \label{fig:DR2_ba_ratios}
\end{figure}

An important characteristic of the CALIFA MS is that it contains galaxies of all morphological types. Galaxy 
morphologies were inferred by combining the independent visual classifications of several collaboration 
members as described in \citetalias{Walcher:2014}. Fig.~\ref{fig:DR2_morph} shows a histogram of bars 
strengths as well as the fraction of DR2 galaxies with respect to the expected final sample distribution 
for different morphological types grouped into elliptical, lenticular and spiral galaxies (and subtypes). 
A more detailed classification of ellipticals (from 0 to 7) is available, but we do not distinguish 
between them here because of the low number of galaxies per elliptical subtype within DR2.
From 200 galaxies in DR2, 18 have been classified as ongoing mergers\footnote{According to our visual 
classification.} (of any type). 
As clearly seen in Fig.~\ref{fig:DR2_morph}, the fraction of DR2 galaxies with respect to the expected 
final sample is almost constant for all types, implying that the DR2 coverage seems to be consistent 
with a random selection. 
Axis ratios (b/a) were measured from the SDSS $r$-band image from growth curve analysis by calculating 
light moments after proper sky subtraction and masking of foreground stars (see \citetalias{Walcher:2014} 
for details). The axis ratios can be used as a proxy of the inclination of spiral galaxies. 
Figure~\ref{fig:DR2_ba_ratios} shows that the DR2 sample covers the same range of axis ratios 
as the final sample. A Kolmogorov-Smirnov test confirms with $>$ 95\% confidence that the morphology 
and the axis-ratio distributions of DR2 are consistent with being randomly drawn from the CALIFA MS.

\begin{figure}
 \resizebox{\hsize}{!}{\includegraphics{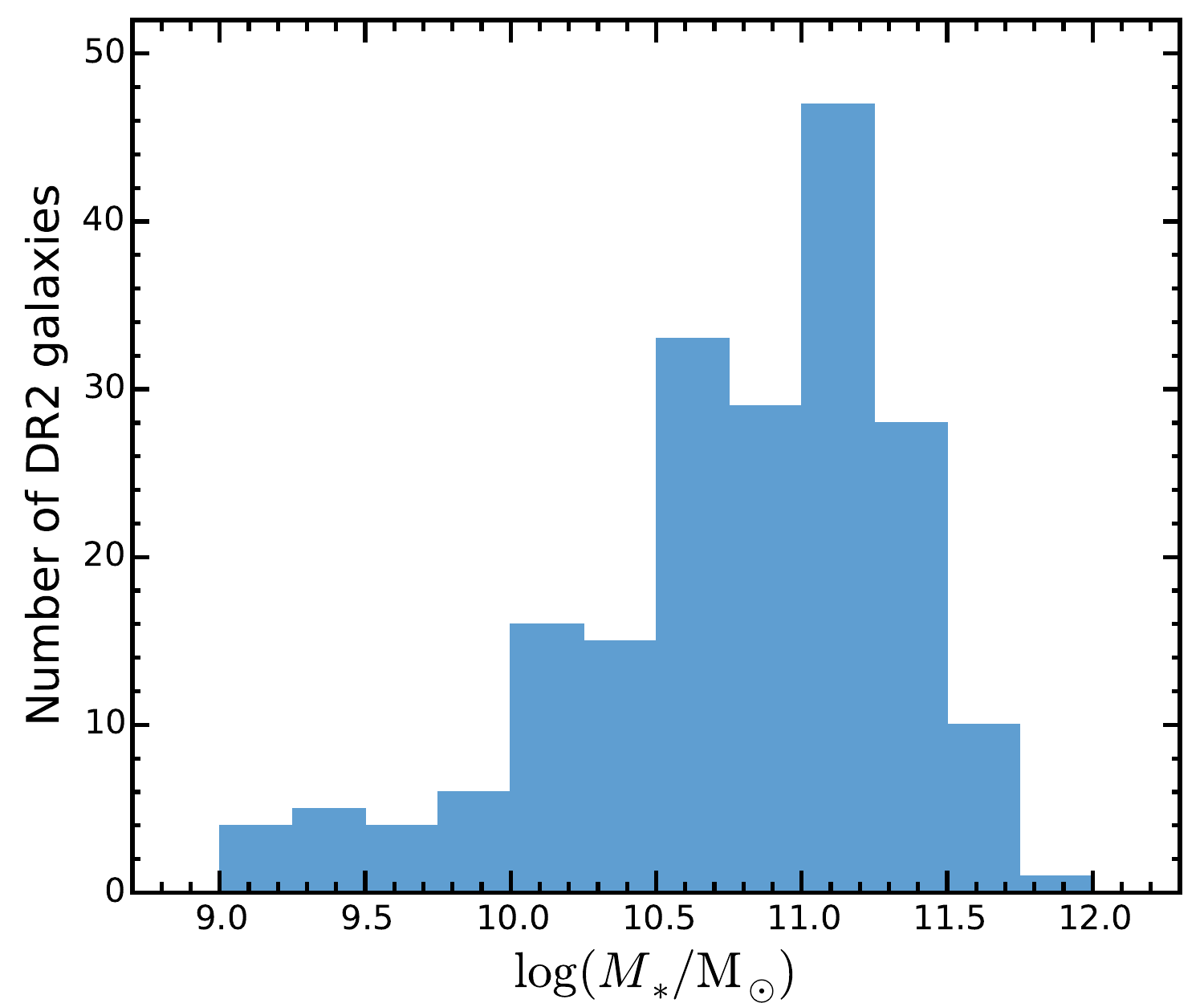}}
  \caption{Distribution of stellar masses in the DR2 sample. The stellar masses have 
	been determined from the CALIFA data using spectral fitting techniques (see text 
	for details).}
  \label{fig:DR2_mass_dist}
\end{figure}

In Fig.~\ref{fig:DR2_mass_dist}, we present the distribution of stellar masses for the DR2 galaxies. 
Galaxy stellar masses are from \citet{GonzalezDelgado:2014b}, and they have been estimated following 
the process described in \citet{Perez:2013}, \citet{CidFernandes:2013,CidFernandes:2014} and 
\citet{GonzalezDelgado:2014a}. These masses account for spatial variations in both M/L 
ratio and stellar extinction.
In short, we use the {\sc starlight} code \citep{CidFernandes:2005} to fit each spectrum extracted 
from the datacube with a combination of SSP models from the Granada \citep{GonzalezDelgado:2005} and 
MILES \citep{Vazdekis:2010} libraries, that cover the full metallicity range of the MILES models 
(log Z/Z$_{\odot}$ from -2.3 to +0.22), and ages from 0.001 to 14 Gyr. We assume a Salpeter IMF.
The DR2 galaxies cover intermediate to high-mass galaxies, including at least 10 galaxies per 0.25\,dex 
bin between $10^{10}$ and $10^{12}M_{\sun}$ and a median value close to $10^{11}\mathrm{M}_{\sun}$. The 
asymmetric distribution is expected from the distribution in absolute magnitudes 
(see Fig.~\ref{fig:DR2_CMD}) and is inherited from the CALIFA MS due to its selection criteria 
(see \citetalias{Walcher:2014} for details).  

\begin{figure*}
 \centering
 \resizebox{18.2cm}{!}{\includegraphics{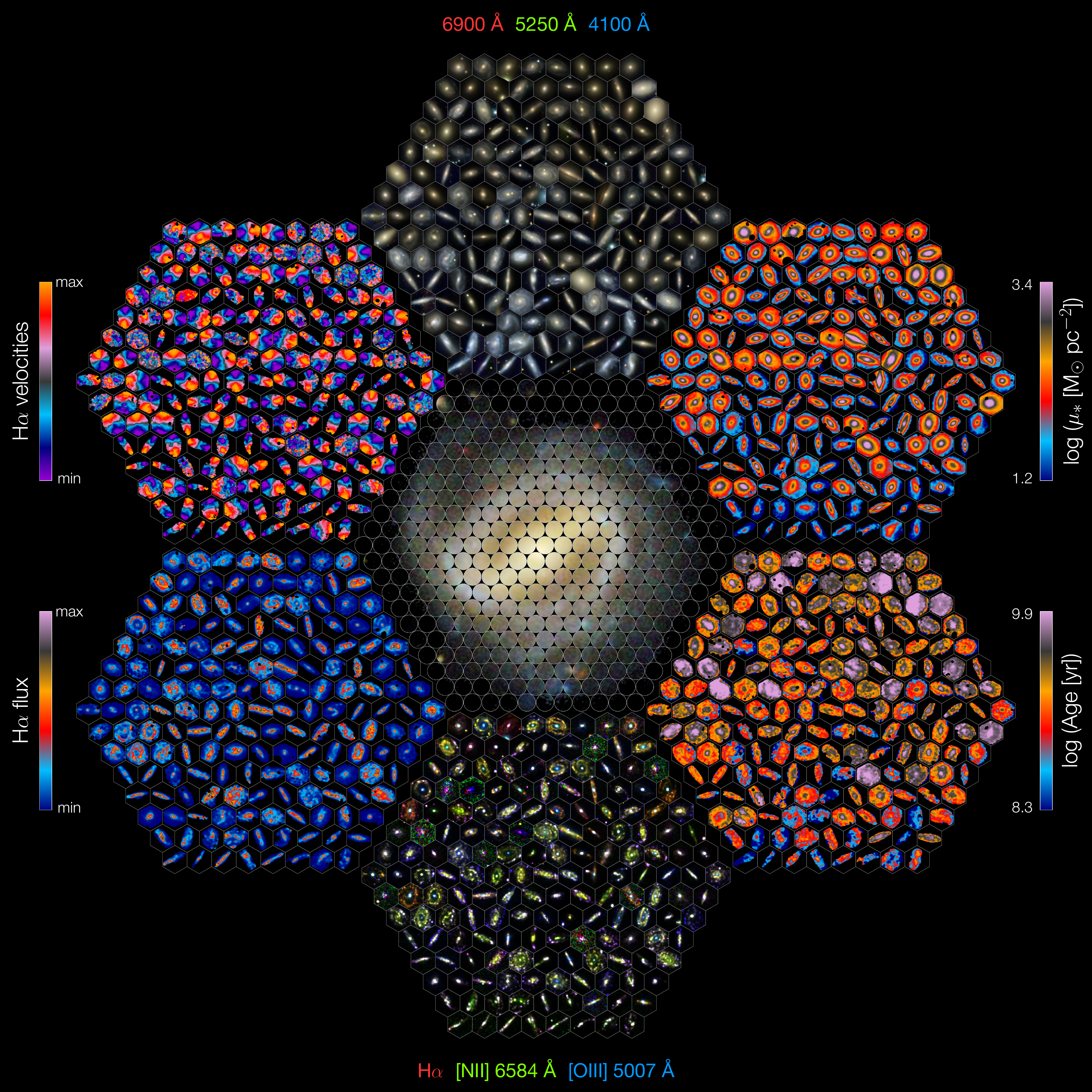}}
  \caption{CALIFA ``\textit{panoramic view}'' (also CALIFA's ``\textit{Mandala}'') representation, 
        consisting of the basic physical properties (all of them derived from the CALIFA datacubes) 
        of a subsample of 169 galaxies extracted randomly from DR2. We show 1) broad band 
        images (top center; central wavelength 6900 \AA, 5250 \AA, and 4100 \AA), 2) stellar mass 
        surface densities (upper right), 3) ages (lower right), 4) narrow band images (bottom center; 
        emission lines: H$\alpha$, [\ion{N}{ii}] 6584 \AA, and [\ion{O}{iii}] 5007 \AA), 5) H$\alpha$ 
        emission (lower left) and 6) H$\alpha$ kinematics (upper left). The CALIFA logo is placed at 
        the central hexagon.}
  \label{fig:DR2Hex}
\end{figure*}

A more general ``\textit{panoramic view}'' of the DR2 sample characteristics is presented in 
Fig.~\ref{fig:DR2Hex}. Several of the main properties observable in 2D are highlighted for 169 
randomly-selected galaxies, shown individually in hexagons that together form the shape of a 
CALIFA-like FoV. 
The galaxies have been ordered by $r$-band absolute magnitude (a proxy for the stellar mass), from 
top right (lowest absolute magnitude) to bottom left (highest absolute magnitude).
The highlighted properties derive from several different analysis pipelines developed within 
the collaboration. Stellar properties like ages and mass surface density were measured with 
the {\sc starlight} code (see references in the preceding paragraph describing the distribution of 
stellar masses in the sample) while gas properties and emission lines were measured using FIT3D
\citep{Sanchez:2007}. This plot is only intended to demonstrate the diversity of the DR2 sample.

\section{Data processing and error propagation}\label{sect:data_processing}

\begin{figure*}
  \resizebox{\hsize}{!}{\includegraphics{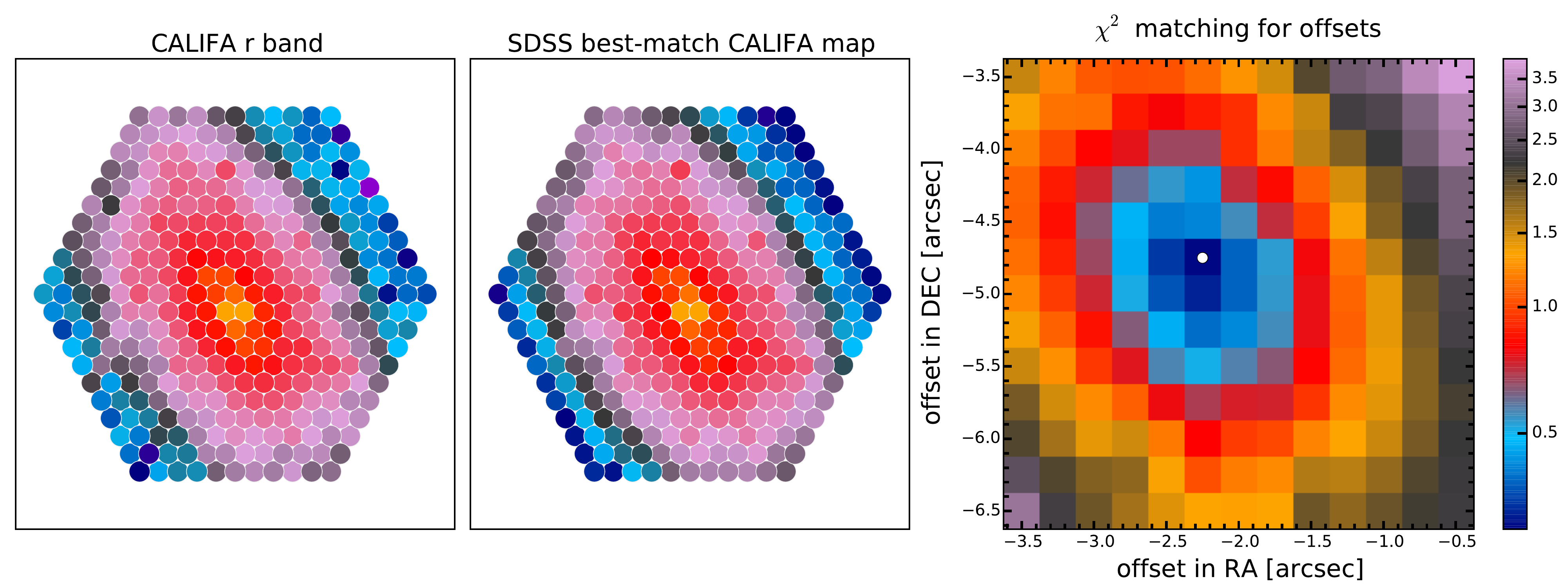}}
  \caption{Example of the registering method for pointing 1 of NGC0496 (ID 45). \emph{Left panel:} Flux map
    in $r$-band for the PPak fibers. \emph{Central panel:} Predicted SDSS flux for each CALIFA fiber estimated 
    using $2.7\arcsec$ diameter apertures and adopting the PPak layout projected on the SDSS image for the 
    best match according to the $\chi^{2}$ map. Note that the PPak layout is not to scale, i.e. relative 
    distances between adjacent fibers have been decreased for the sake of clarity. 
    \emph{Right panel:} $\chi^{2}$ map of the offsets (best offset marked with a white dot).}
  \label{fig:DR2_reg_chi}
\end{figure*}

For the sake of completeness, we provide here a brief summary of the instrument layout and observing 
strategy. All the details can be found in \citetalias{Sanchez:2012a}.
The PPak fiber bundle of the PMAS instrument has a FoV of 74\arcsec\ $\times$ 64\arcsec. 
There are 382 fibers in total, distributed in 3 different groups. The PPak Integral Field Unit (IFU) 
holds 331 ``science'' fibers in a hexagonal grid with a maximum diameter of 74\arcsec\ while each fiber 
projects to 2\farcs7 in diameter on the sky. The fiber-to-fiber distance is 3\farcs2 which yields a total 
filling factor of 0.6. An additional set of 36 fibers devoted to measuring the surrounding sky level 
are distributed in six bundles of 6 fibers each, located in a circle 72\arcsec\ from the center. 
Finally, there are 15 extra fibers connected to the calibration unit.

Every galaxy in the CALIFA sample is observed in the optical range using two different overlapping 
setups. The V500 low-resolution mode (R $\sim$ 850) covers the range 3745-7500 \AA, but it is 
affected by internal vignetting within the spectrograph giving an unvignetted range of 4240-7140 \AA. 
The blue mid-resolution setup (V1200; R $\sim$ 1650) covers the range 3400-4840 \AA\, with an 
unvignetted range of 3650-4620 \AA. The resolutions quoted are those at the overlapping wavelength 
range ($\lambda$ $\sim$ 4500 \AA). In order to reach a filling factor of 100\% across the FoV, a 
3-pointing dithering scheme is used for each object. The exposure time per pointing is fixed. 
V1200 observations are carried out during dark nights with an exposure time of 1800 s 
(split in 2 or 3 individual exposures) pointing. V500 observations are taken during grey nights 
with 900 s per pointing.

In the following section we describe the new improvements to the CALIFA data reduction pipeline 
used to produce the DR2 data. 

\subsection{Improvements on the CALIFA data reduction scheme}\label{sect:pipeline}

As described in \citetalias{Husemann:2013}, since V1.3c the CALIFA pipeline has a 
{\tt Python}-based architecture ({\tt Py3D} package). The main improvements to the current pipeline 
V1.5 are: i) new sensitivity curve for V500 setup obtained from a dedicated calibration programme for 
several CALIFA elliptical galaxies (Husemann et al., in preparation) ii) a new registering method, 
comparing individual CALIFA pointings with SDSS images; iii) an improved image reconstruction method 
(cube interpolation). Among others, step ii) also improves the absolute photometric matching of the 
three dithered pointings.

The new version starts with the Raw Stacked Spectra (RSS) files of the three individual pointings after 
sky subtraction produced by pipeline V1.3c. The V500 RSS files are then spectrophotometrically re-calibrated 
with the new sensitivity curve (undoing the V1.3c calibration). A new estimate for the sensitivity curve was 
necessary to account for severe wavelength-dependent aperture losses particular to standard star observations 
that have low fiber filling factor given the large fiber diameter of PPak. We therefore re-observed about 
two dozen elliptical CALIFA galaxies with the PMAS Lens-Array (LArr) and the V300 grism with a continuous 
$16\arcsec\times16\arcsec$ FoV covering the bright center of the galaxy. The details of those observations 
and their application to coarse-fiber IFS will be presented in a separate publication (Husemann et al., 
in preparation), but we briefly outline the concept and its application to CALIFA data here. 

A robust spectrophotometric calibration can be assembled using observations of standard stars in the 
PMAS LArr mode, in which aperture losses are absent, and the atmospheric extinction curve has been directly 
estimated for each observing night, instead of using the average extinction curve derived by 
\citet{Sanchez:2007}. We derive a new sensitivity curve using these secondary spectrophotometric standards, 
by comparing the LArr flux spectra against the observed count spectra in targeted elliptical galaxies over 
the same aperture. The latter should be much less sensitive to wavelenght-dependent aperture losses, given 
that the surface brightness profiles of elliptical galaxies smoothly vary over several tens of arcseconds. 
To match the apertures and increase the S/N, all CALIFA fibers that fall within the LArr FoV are co-added 
together. PPak spectra are further smoothed to 9\AA\ FWHM, matching the spectral resolution of the V300 
LArr observations, to improve the match betwen the spectra. To the resulting sensitivity curve we 
fit a high-order polynomical, creating a noise-free representation. Then we derive a master sensitivity 
curve by averaging the sensitivity curves measured independently for each galaxy. We anticipate that 
the largest uncertainty in the relative spectrophotometry across the CALIFA wavelength range will be 
dominated by the unknown extinction curve at the time of each observation. 

The current pipeline also implements a new scheme for estimating the registration of the images. First, 
sky-subtracted and calibrated images are created from SDSS DR7 (in the $r$-band for the V500 setup and 
the $g$-band for the V1200) based on the so-called “corrected frames” ({\tt fpC}). Then, the magnitude 
in the corresponding SDSS filter is computed for each RSS spectrum. The predicted SDSS flux for each 
CALIFA fiber is estimated using $2\farcs7$ diameter apertures, adopting the PPak layout projected on the 
SDSS image. This layout is displaced in steps in RA and Dec across a search box in the SDSS image. Then a 
$\chi^{2}$ map is computed to obtain the best offsets for each pointing, taking into account errors in the 
flux measurements (only fibers with S/N > 3.0 are considered) and allowing for a photometric scaling factor 
between the SDSS and the CALIFA observations as an additional parameter. The minimum value of the $\chi^{2}$ 
map is used to obtain the best-fitting RA and Dec for the center of the PPak IFU with respect to the center 
of the CALIFA galaxy seen by SDSS. Figure~\ref{fig:DR2_reg_chi} shows an example of the described procedure. 
The photometric scale factor at the best matching position is used to rescale the absolute photometry of 
each particular RSS pointing to bring them on the same flux scale. 

The photometric anchoring to the SDSS images of the V1.5 data is more accurate than those of the previous 
version. However, there are a few datacubes where the new registering method does not return 
optimal results, particularly in low surface brightness edge-on galaxies or in the presence 
of bright foreground field stars. This effect is more likely to occur in the V1200 setup, given its lower 
S/N on average compared to V500. In such cases, we apply the photometric SDSS matching of pipeline 
V1.3c described in \citetalias{Husemann:2013} (to both setups, for the sake of consistency). A new 
``REGISTER'' keyword has been included in the header of the datacubes (see Sect.~\ref{sect:fheader}) and a 
dagger symbol has been added to the quality tables (Table \ref{tab:QC_par_V500} and \ref{tab:QC_par_V1200}) 
in order to easily identify these galaxies.

The third step in the reduction sequence is the interpolation method used to convert from RSS to cube format,
aimed at improving the spatial resolution. We use the position of each RSS pointing obtained in the previous 
step for the image reconstruction. In a series of tests, we found that an inverse-distance weighted image 
reconstruction scheme performs more favorably than, e.g. the drizzle method \citep{Fruchter:2002}. In order 
to increase the spatial resolution and reduce the correlation between nearby pixels, we have reduced the 
extent of the Gaussian kernel for the interpolation. We adopt 0.75\arcsec\ for the dispersion of the Gaussian 
(instead of 1\arcsec\ in V1.3c) and limit the kernel to a radius of 3.5\arcsec\ (instead of 5\arcsec).
This results in a much sharper image and a lower value for the correlated noise. In the previous pipeline V1.3c, 
a minimum number of 3 fibers was imposed in the reconstruction of the image to achieve a homogeneous data 
quality across the field. With the new maximum radius in pipeline V1.5, such prescription results in the 
absence of data in the outer 2$\arcsec$ of the FoV, due to the wider fiber separation in the outer ring 
of the fiber bundle. Thus, we decided to lower this limit to 1 as the minimum number of fibers needed to 
fill a spaxel. We have added a new Header Data Unit (see Sect.~\ref{sect:fcover}) that records the number 
of fibers used to compute the total flux per spaxel. This allows the user to control what spaxels to include 
if a particular science case requires a minimum number of fibers for the reconstruction of the flux.

\subsection{Characterization of spatially correlated noise}\label{sect:correlation}

Due to the interpolation procedure to obtain a regular grid, the output pixels in the final datacube are 
not independent of one another. The gaussian interpolation method distributes the flux from a given fiber 
between several pixels which are combined with neighboring pixels within a certain radius, as described 
in Sect.~\ref{sect:pipeline}. This causes the noise in the adjacent pixels to be correlated (in the 
spatial dimension). The correlation implies that a measurement of the noise in a stacked spectrum of 
$N$ pixels will be underestimated (noise is underestimated on scales larger than pixel units). 
Characterizing this effect is essential for estimating the statistical errors when spectra in datacubes 
are co-added to increase the S/N, a common approach in specific applications when a minimum S/N is 
required. 

\begin{figure}
  \resizebox{\hsize}{!}{\includegraphics{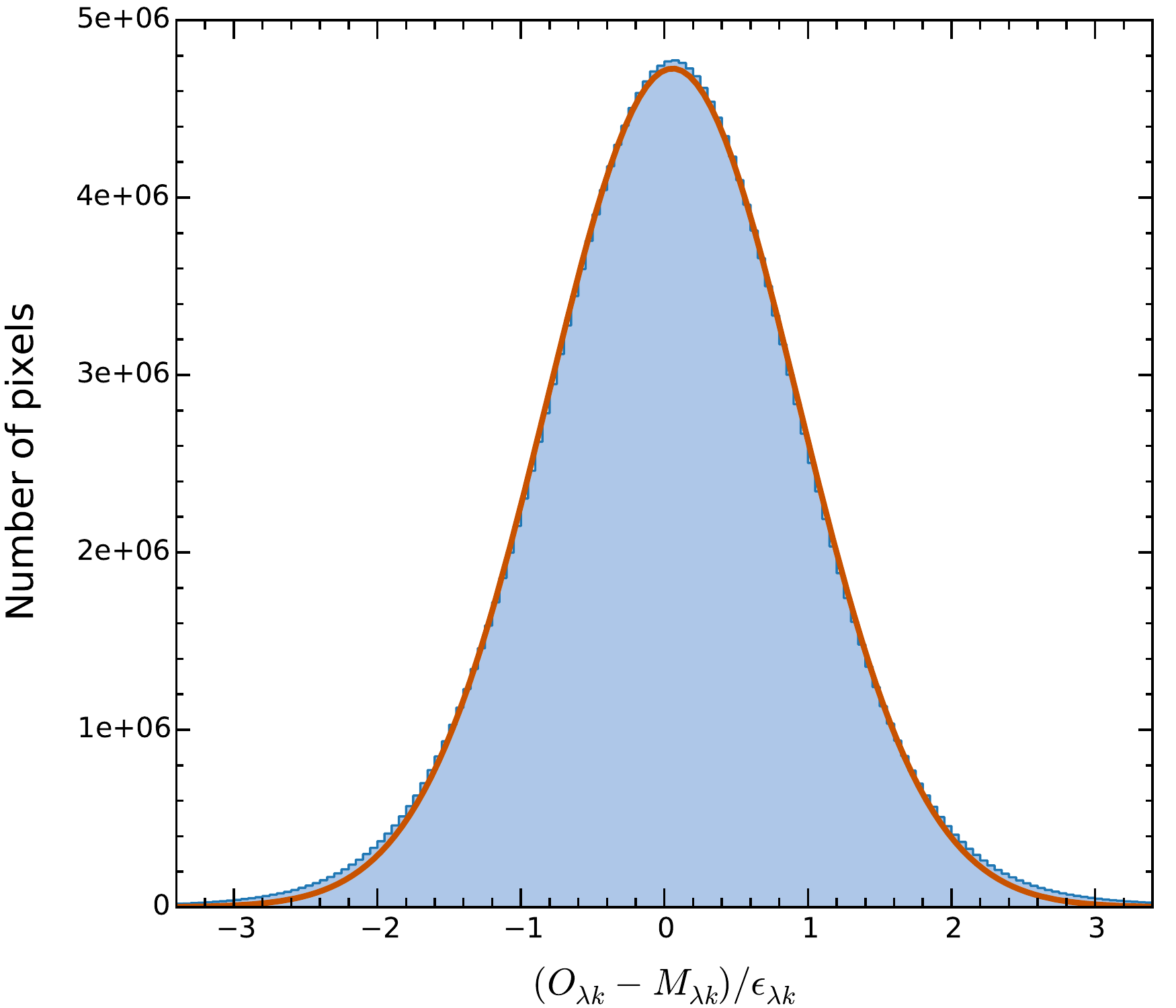}}
  \caption{Histogram of the reduced residuals $(O_{\lambda,k} - M_{\lambda,k}) / \epsilon_{\lambda,k}$ for 
     all $\lambda$'s, all bins ($k$) and all galaxies in DR2 (209151086 points in total). The solid orange 
     line shows the best Gaussian fit to the sample.}
  \label{fig:DR2_hist_error}
\end{figure}

First of all, it is important to check that the error spectra derived from the pipeline for individual 
spaxels are reliable. Spectral fitting analysis can provide an approximate assessment of the accuracy 
of the error spectra. In Fig.~\ref{fig:DR2_hist_error} we update figure~9 of \citetalias{Husemann:2013} to 
DR2 data. The plot shows the histogram of reduced residuals, i.e. the difference between the observed 
($O_{\lambda}$) and synthetic ($M_{\lambda}$) spectra obtained with {\sc starlight} in units of the 
corresponding error $\epsilon_{\lambda}$ (details on the fitting procedures can be found in 
\ref{sect:specphot_cal}). The distribution is very well described by a Gaussian centered at 0.03 with 
$\sigma = 0.87$, only slightly less than expected if residuals are purely due to noise. 

\begin{figure}
  \resizebox{\hsize}{!}{\includegraphics{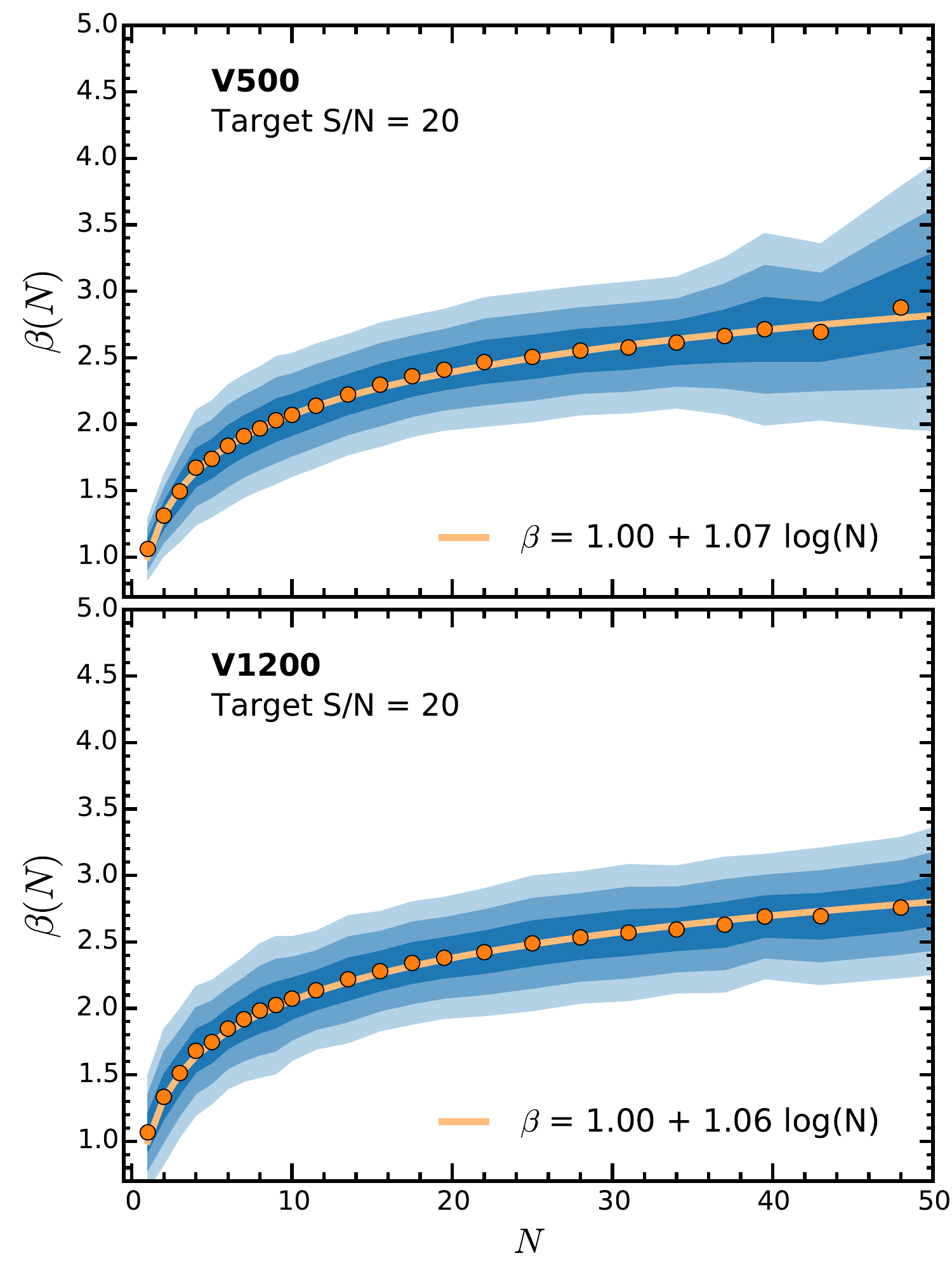}}
  \caption{Noise correlation ratio $\beta$ (ratio of the real estimated error to the analytically 
    propagated error) as a function of number of spaxels per bin for all the V500 (upper panel) 
    and V1200 (lower panel) data of DR2 at a target S/N of 20. Shaded areas mark the 1$\sigma$, 2$\sigma$ 
    and 3$\sigma$ levels. The orange lines represent the best fit logarithmic function with a slope 
    $\alpha = 1.07$ and $\alpha = 1.06$, respectively.}
  \label{fig:DR2_noise_correlation}
\end{figure}

The correlated noise can be taken into account by providing the spatial covariance \citep{Sharp:2014}. 
However, a more practical approach consists of using the datacubes to calculate the expected rms noise,  
with the noise correlation ratio $\beta (N)$, as a function of the number of pixels. To obtain a sample of 
co-added spaxels with different areas, we have used the Voronoi adaptive binning method \citep[implemented
for optical IFS data by][]{Cappellari:2003} with a target S/N of 20. We have removed from the analysis 
individual spaxels with S/N $<$ 5 and co-added bins with areas larger than 60 spaxels. The $\beta$ 
correlation ratio (or correction factor) is the ratio of the ``real'' or measured error to the 
analytically propagated error of the binned spectra as a function of bin size. The results obtained 
for all DR2 datacubes, shown in Fig.~\ref{fig:DR2_noise_correlation}, can be well 
described by the logarithmic function

\begin{equation}
 \beta (N) = 1+\alpha\log N,\label{eq:correlation}
\end{equation}
with $N$ the number of spaxels per bin.

The values for the slope $\alpha$ are equal within the errors (0.01) in both setups, with a value of 1.06 
for V1200 and 1.07 for V500. The slope is lower than the DR1 value (mean $\sim$ 1.4), indicating that 
the noise in DR2 datacubes is less correlated than in DR1. This is expected since we changed the 
parameters in the interpolation (reducing the number of adjacent fibers contributing to a particular 
spaxel) and the registering method. In Appendix~\ref{ap:correlation} we give some instructions on 
how to estimate the final co-added error spectrum and the limit of the application of equation 
\ref{eq:correlation}.


\begin{table*}
\centering
\caption{CALIFA FITS file structure}
\label{tab:HDUs}
\begin{tabular}{cccc}\hline\hline
HDU & Extension name & Format & Content\\\hline
0 & Primary & 32-bit float & flux density in units of $10^{-16}\,\mathrm{erg}\,\mathrm{s}^{-1}\,\mathrm{cm}^{-2}\,\mathrm{\AA}^{-1}$\\
1 & ERROR & 32-bit float & $1\sigma$ error on the flux density\\
2 & ERRWEIGHT & 32-bit float & error weighting factor\\
3 & BADPIX &   8-bit integer & bad pixel flags (1=bad, 0=good) \\
4 & FIBCOVER &  8-bit integer & number of fibers used to fill each spaxel\\ \hline
\end{tabular}
\end{table*}

\section{CALIFA data format and characteristics}\label{sect:data_format}
The CALIFA data are stored and distributed as datacubes (three-dimensional data) in the standard 
binary FITS format and consist of several FITS Header Data Units (HDU). These datacubes contain 
(1) the measured flux densities, corrected for Galactic extinction as described in 
\citetalias{Sanchez:2012a}, in units of 
$10^{-16}\,\mathrm{erg}\,\mathrm{s}^{-1}\,\mathrm{cm}^{-2}\,\mathrm{\AA}^{-1}$ (primary datacube), (2) 
associated errors, (3) error weighting factors, (4) bad pixels flags and (5) fiber coverage 
(Table~\ref{tab:HDUs}). The last HDU is a new added content absent in DR1, as explained in 
Sect.~\ref{sect:pipeline}, but the others share the same properties as the previous data release. 
The first two axes of the cubes correspond to the spatial dimension along right ascension and 
declination with a $1\arcsec\times1\arcsec$ sampling. The third dimension represents the wavelength 
and is linearly sampled. Table~\ref{tab:cube_dimension} summarizes the dimensions of each datacube 
($N_\alpha$, $N_\delta$, and $N_\lambda$), as well as the spectral sampling ($d_\lambda$) and constant 
resolution ($\delta_\lambda$) along the entire wavelength range.

\subsection{Error and weight datacubes}
The $1\sigma$ noise level of each pixel as formally propagated by the pipeline
can be found in the first FITS extension. Sect.~\ref{sect:correlation} discusses on the accuracy 
of the formal noise and the correlation, important when CALIFA data need to be spatially binned, 
and an empirical function is provided to account for the correlation effect. The second FITS 
extension (ERRWEIGHT) stores the error scaling factor for each pixel in the limiting case 
that all valid spaxels of the cube would be co-added (see also Appendix \ref{ap:correlation}).
In the case of bad pixels, we assigned an error value that is roughly ten orders of magnitude higher 
than the typical value. 

\begin{table}
\caption{Dimension and sampling of CALIFA datacubes}
\label{tab:cube_dimension}
\begin{tabular}{lccccccc}\hline\hline
\small{Setup} & \small{$N_\alpha$}\tablefootmark{a} & \small{$N_\delta$}\tablefootmark{a} & \small{$N_\lambda$}\tablefootmark{a} & \small{$\lambda_\mathrm{start}$}\tablefootmark{b} & \small{$\lambda_\mathrm{end}$}\tablefootmark{c} & \small{$d_\lambda$}\tablefootmark{d} & \small{$\delta_\lambda$}\tablefootmark{e} \\\hline
\small{V500}  &  \small{78}    & \small{73}    &  \small{1877}   & \small{3749\AA}     & \small{7501\AA}  & \small{2.0\AA}  & \small{6.0\AA}    \\
\small{V1200} &  \small{78} & \small{73} &  \small{1701} & \small{3650\AA} & \small{4840\AA} & \small{0.7\AA} & \small{2.3\AA}  \\\hline
\end{tabular}
\tablefoot{\tablefoottext{a}{Number of pixels in each dimension.}
\tablefoottext{b}{Wavelength of the first pixel on the wavelength direction.}
\tablefoottext{c}{Wavelength of the last pixel on the wavelength direction.}
\tablefoottext{d}{Wavelength sampling per pixel.}
\tablefoottext{e}{Homogenized spectral resolution (FWHM) over the entire wavelength range.}
}
\end{table}

\subsection{Bad pixel datacubes}
Bad pixel datacubes are stored in the third FITS extension (BADPIX). This information, in combination with the 
error vector, is essential to properly account for the potential problems in each spaxel. Pixels 
with flag = 1 reports the absence of sufficient information in the raw data due to cosmic rays, bad CCD colums or 
the effect of vignetting\footnote{The vignetting effect imprints a characteristic inhomogeneous pattern across the 
FoV on the bad pixels vector. See Fig.~11 of \citetalias{Husemann:2013} for more details.}.
These bad pixels have been interpolated and we strongly suggest not to use them for any science analysis. 

Finally, the uncovered corners of the hexagonal PPak FoV are filled with zeros and flagged as bad pixels for 
consistency. The residuals of bright night-sky emission lines are not flagged as bad pixels. 

\subsection{Fiber coverage datacubes}\label{sect:fcover}

Pipeline V1.5 adds a new FITS extension (FIBCOVER) to the datacubes, not available in previous DR1 datacubes. 
As explained in Sect.~\ref{sect:pipeline} we have reduced the maximum distance of fibers that can contribute 
to the flux of a given spaxel. The outer hexagonal-ring of fibers do not have the same coverage in the 
surroundings as any other fiber inside the hexagon. In pipeline V1.3c we imposed a minimum of 3 fibers for 
computing the flux of given spaxel. In V1.5, with the new radius limit this would yield an empty outer 
hexagonal-ring of $\sim$ 2\arcsec\ in the FoV. Thus, we have relaxed to 1 the minimum number of fibers. In 
order to control which spaxels have enough flux ``resolution'', we have included a new HDU reporting the 
number of fibers used to account for the computed flux.

\subsection{FITS header information}\label{sect:fheader}

The FITS header contains the standard keywords that encode the information required to transform the 
pixel-space coordinates into sky and wavelength-space coordinates, following the 
World Coordinate System \citep[WCS,][]{Greisen:2002}.
Each CALIFA datacube contains the full FITS header information of all raw frames from which it was created. 
Information regarding observing and instrumental conditions such us sky brightness, flexure 
offsets, Galactic extinction or approximate limiting magnitude is also kept in the FITS header of each 
datacube. See Sect.~4.3 of \citetalias{Husemann:2013} for nomenclature and their Table 4 for a summary of the 
main header keywords and meaning. 

The most important new keyword added in DR2 datacubes is ``REGISTER'' and takes a boolean value. It indicates 
if a particular datacube has been successfully registered using the new method explained in 
Sect.~\ref{sect:pipeline} (\emph{True}) or it has used the old V1.3c scheme (\emph{False}). 
Datacubes with a \emph{False} value are marked with a dagger in Tables \ref{tab:QC_par_V500} 
and \ref{tab:QC_par_V1200}.


\section{Data Quality}\label{sect:QC}

This second CALIFA data release (DR2) provides science-grade data for a sample of 200 galaxies, including the 
100 galaxies released in the first data release (DR1), identified by an asterisk in Tables \ref{tab:QC_par_V500} 
and \ref{tab:QC_par_V1200}. As for DR1, we have run a careful quality control (QC) on the data products and 
selected only those galaxies that passed a series of QC checks in both setups 
(V500 and V1200), as we detail in this section.
The QC checks are based on a set of measured parameters and/or visual inspection, resulting in a set of flags that 
allow to quickly assess the quality of the data and their suitability for scientific use.
Quantities and flags are organized into three distinct categories, respectively related to:
observing conditions (denoted by the \textsc{obs} prefix); instrumental performance and effectiveness of the data 
reduction (\textsc{red}); accuracy and quality of the final data products (\textsc{cal}). The flags in each category 
are computed based on thresholds on measured quantities, possibly combined with flags given by human classifiers based 
on visual inspection, as detailed below and summarized in Tables \ref{tab:QCflags_def_V500} 
and \ref{tab:QCflags_def_V1200}. Thresholds are determined from the distribution of the parameters in order to 
exclude outliers and also by analyzing the effects of anomalous parameters on the final quality of the datacubes.
The tables of the relevant QC parameters, along with the QC flags are available on the DR2 website.

Each flag can have one of the following values:

\begin{description}
\renewcommand{\labelitemi}{$\bullet$}
\item[$\bullet$] $-1 =$~ undefined
\item[$\bullet$] $~0 =$~ good quality -- \textsc{OK}
\item[$\bullet$] $~1 =$~ minor issues that do not significantly affect the quality -- \textsc{warning}
\item[$\bullet$] $~2 =$~ significant issues affecting the quality -- \textsc{bad}
\end{description}

By selection, DR2 only includes galaxies with \textsc{warning} flags in the worst cases, with just a few minor 
exceptions affecting previously released DR1 galaxies: in these cases the revised QC criteria adopted here would 
have prevented to include such galaxies in the DR, but given the incremental nature of our data releases we keep 
them in the current sample.

In naming the QC parameters we adopt the following convention: the first part is the category prefix (\textsc{obs}, 
\textsc{red} or \textsc{cal}), followed by a measured parameter, and possibly a final suffix indicating the 
statistics applied to combine the parameter as measured in different observations/pointings/fibers 
(i.e., \textsc{mean}, \textsc{min}, \textsc{max}, \textsc{rms}).

\begin{table*}
\centering
\caption{Definition of CALIFA DR2 quality control flags for the V500 data }
\label{tab:QCflags_def_V500}
\begin{tabular}{lllll}
\hline\hline\\
QC flag & QC parameters involved & \textsc{warning} condition(s) & \textsc{bad} condition(s) & Flag definition \\
\hline\hline\\
\textsc{flag\_obs\_am} & \textsc{obs\_airmass\_mean} & $>1.7$ & $>2.0$ & Worst of the three parameters \\
                       & \textsc{obs\_airmass\_max}  & $>2.0$ & $>2.5$ & \\
                       & \textsc{obs\_airmass\_rms}  & $>0.15$ & ... & \\
\hline\\
\textsc{flag\_obs\_skymag} & \textsc{obs\_skymag\_mean} & $<20.5\,\mathrm{mag_V\,arcsec}^{-2}$ & $<19.5$ & Worst of the two parameters \\
                           & \textsc{obs\_skymag\_rms}  & $>0.1$ & ... & \\
\hline\\
\textsc{flag\_obs\_ext} & \textsc{obs\_ext\_mean} & $>0.30$~mag & ... & Worst of the three parameters \\
                        & \textsc{obs\_ext\_max}  & $>0.35$     & ... & \\
                        & \textsc{obs\_ext\_rms}  & $>0.10$     & ... & \\
\hline
\hline\\
\textsc{flag\_red\_straylight} & \textsc{red\_meanstraylight\_max} & $>30$ counts& $>50$ & Worst of the three parameters \\
                        & \textsc{red\_maxstraylight\_max} & $>50$ & $>100$ &  \\
                        & \textsc{red\_rmsstraylight\_max} & $>5$ & $>10$ &  \\
\hline\\
\textsc{flag\_red\_disp} & \textsc{red\_disp\_mean} & $>5.5$ \AA~(FWHM) & ... & Worst of the three parameters \\
                        & \textsc{red\_disp\_max} & $>10.0$ & ... &  \\
                        & \textsc{red\_disp\_rms} & $>0.5$ & $>1.0$ &  \\
\hline\\
\textsc{flag\_red\_cdisp} & \textsc{red\_cdisp\_mean} & $>3.0$ pixels (FWHM) & ... & Worst of the three parameters \\
                        & \textsc{red\_cdisp\_max} & $\ge 4.0$ & ... &  \\
                        & \textsc{red\_cdisp\_rms} & $>0.25$ & ... &  \\
\hline\\
\textsc{flag\_red\_skylines} & \textsc{red\_res5577\_min} & $<-0.1$ counts & ... & Worst of the three parameters \\
                             & \textsc{red\_res5577\_max} & $>0.1$ & ... &  \\
                             & \textsc{red\_rmsres5577\_max} & $>1.0$ & ... &  \\
\hline\\
\textsc{flag\_red\_limsb} & \textsc{red\_limsb} & $<23.25\,\mathrm{mag_V\,arcsec}^{-2}$ & $<22.50$ & \\
\hline
\hline\\
\textsc{flag\_cal\_specphoto} & \textsc{cal\_qflux\_g} & $>0.06$ dex & $>0.097$ dex & Worst of the three parameters \\
                              &                        & $<-0.06$ dex& $<-0.097$ dex& combined with visual checks \\
                              & \textsc{cal\_qflux\_r} & $>0.06$ dex & $>0.097$ dex & on the 30"-integrated spectrum: \\
                              &                        & $<-0.06$ dex& $<-0.097$ dex& spectral shape and comparison \\
                              & \textsc{cal\_qflux\_rms} & $>0.1$ & $>0.2$ & with SDSS photometry \\
\hline\\
\textsc{flag\_cal\_wl} & \textsc{cal\_rmsvelmean} & $>2.0~\mathrm{km~s}^{-1}$  & $>5.0$ & \\
\hline\\
\textsc{flag\_cal\_ima} & \textsc{cal\_chi2reg\_max} & $>10$  & ... & Combine parameter and visual\\
&&&& inspection on registration and\\
&&&& synthetic broad-band image\\
\hline\hline\\
\end{tabular}
\end{table*}

\begin{table*}
\centering
\caption{Definition of CALIFA DR2 quality control flags for the V1200 data }
\label{tab:QCflags_def_V1200}
\begin{tabular}{lllll}
\hline\hline\\
QC flag & QC parameters involved & \textsc{warning} condition(s) & \textsc{bad} condition(s) & Flag definition \\
\hline\hline\\
\textsc{flag\_obs\_am} & \textsc{obs\_airmass\_mean} & $>1.7$ & $>2.0$ & Worst of the three parameters \\
                       & \textsc{obs\_airmass\_max}  & $>2.0$ & $>2.5$ & \\
                       & \textsc{obs\_airmass\_rms}  & $>0.15$ & ... & \\
\hline\\
\textsc{flag\_obs\_skymag} & \textsc{obs\_skymag\_mean} & $<21.5\,\mathrm{mag_V\,arcsec}^{-2}$ & $<21.0$ & Worst of the two parameters \\
                           & \textsc{obs\_skymag\_rms}  & $>0.1$ & ... & \\
\hline\\
\textsc{flag\_obs\_ext} & \textsc{obs\_ext\_mean} & $>0.30$~mag & ... & Worst of the three parameters \\
                        & \textsc{obs\_ext\_max}  & $>0.35$     & ... & \\
                        & \textsc{obs\_ext\_rms}  & $>0.10$     & ... & \\
\hline
\hline\\
\textsc{flag\_red\_straylight} & \textsc{red\_meanstraylight\_max} & $>15$ counts& $>30$ & Worst of the three parameters \\
                        & \textsc{red\_maxstraylight\_max} & $>20$ & $>40$ &  \\
                        & \textsc{red\_rmsstraylight\_max} & $>1.5$ & $>2.0$ &  \\
\hline\\
\textsc{flag\_red\_disp} & \textsc{red\_disp\_mean} & $>2.0$ \AA~(FWHM) & $>2.5$ & Worst of the three parameters \\
                        & \textsc{red\_disp\_max} & $>10.0$ & ... &  \\
                        & \textsc{red\_disp\_rms} & $>0.15$ & $ ... $ &  \\
\hline\\
\textsc{flag\_red\_cdisp} & \textsc{red\_cdisp\_mean} & $>3.0$ pixels (FWHM) & ... & Worst of the two parameters \\
                        & \textsc{red\_cdisp\_rms} & $>0.66$ & ... &  \\
\hline\\
\textsc{flag\_red\_skylines} & \textsc{red\_res4358\_min} & $<-0.1$ counts & ... & Worst of the three parameters \\
                             & \textsc{red\_res4358\_max} & $>0.1$ & ... &  \\
                             & \textsc{red\_rmsres4358\_max} & $>0.7$ & ... &  \\
\hline\\
\textsc{flag\_red\_limsb} & \textsc{red\_limsb} & $<22.50\,\mathrm{mag_B\,arcsec}^{-2}$ & $<22.00$ & \\
\hline
\hline\\
\textsc{flag\_cal\_specphoto} &  &  &  & Visual checks on 30"-aperture \\
                              &  &  &  & integrated spectrum for \\
                              &  &  &  & spectral shape and mismatch with \\
                              &  &  &  & V500 spectrophotometry \\
\hline\\
\textsc{flag\_cal\_wl} & \textsc{cal\_rmsvelmean} & $>1.0~\mathrm{km~s}^{-1}$  & $>2.0$ & \\
\hline\\
\textsc{flag\_cal\_ima} & \textsc{cal\_chi2reg\_max} & $>10$  & ... & Combined parameter of visual\\
&&&& inspection on registration and\\
&&&& synthetic broad-band image\\
\hline\hline\\
\end{tabular}
\end{table*}

In the following subsections we describe the QCs in each of the above-mentioned categories. As mentioned in 
Sect.~\ref{sect:pipeline}, the V1.5 pipeline starts after sky subtraction of the individual RSS files. Thus, 
some of the quality and properties of the DR2 datacubes are inherited from V1.3c and will not be discussed 
here, namely: wavelength calibration and sky subtraction.

\subsection{Quality of the observing conditions (\textsc{obs})}\label{subsect:QC_obs}

Three quantities are considered crucial in determining the quality of the observing conditions of the CALIFA data: 
the airmass, the brightness of the sky, and the atmospheric extinction. While seeing is in general an important 
parameter of the observing conditions, the imaging quality and spatial resolution of the CALIFA cubes is mostly 
limited by the sampling of the fibers on the plane of the sky and the resampling process 
(see section \ref{sect:astrometry} for more detail), rather than by the seeing. Moreover, the seeing 
measurement is only available for a small fraction of the objects (see Sect.~\ref{sect:spatial_resolution}), and 
therefore cannot be used as a reliable QC parameter.

For the airmass we consider the average and the maximum airmass of the observations over all pointings 
(\textsc{obs\_airmass\_mean} and \textsc{obs\_airmass\_max}) and its rms (\textsc{obs\_airmass\_rms}). For 
each of these quantities we defined two thresholds (the same for V500 and V1200, see Table \ref{tab:QC_par_V500} 
and \ref{tab:QC_par_V1200}) above which the \textsc{warning} or the \textsc{bad} flags, respectively, are raised. 
The combined \textsc{flag\_obs\_am} is the worst of the three cases.

The surface brightness of the sky in V-band during the observations is another critical parameter, which mainly 
limits the depth of the observations and the accuracy of the sky subtraction. The quantity \textsc{skymag} is 
measured in each pointing from the sky spectrum obtained from the 36 sky fibers\footnote{See Appendix A.8 
of \citetalias{Husemann:2013}.}. The mean and the rms over all pointings are considered to define the 
corresponding flags. Note that stricter requirements are applied to V1200 data (blue setup, high resolution) with 
respect to the V500 ones.

The transparency of the sky during each pointing (\textsc{ext}) is obtained from the monitored $V$ band extinction 
at the time of the observation. We consider as symptoms of low/bad quality observations large extinctions on average, 
a large maximum extinction or a large rms variation across the pointings (indicating inhomogeneous observing conditions).

\begin{longtab}
\begin{longtable}{lcccccccccccc}
\caption{CALIFA DR2 quality control parameters for the V500 data }\\\hline\hline
\label{tab:QC_par_V500}
\small{ID\tablefootmark{a}} & \small{airmass\tablefootmark{b}} & \small{$\mu_{V,\mathrm{sky}}$\tablefootmark{c}} & \small{$A_V$\tablefootmark{d}} & \small{seeing\tablefootmark{e}} & \small{flags(O)\tablefootmark{f}} & \small{$\delta_\lambda$\tablefootmark{g}} & \small{S/N($R_{50}$)\tablefootmark{h}} & \small{$I_{3\sigma}$\tablefootmark{i}} & \small{flags(R)\tablefootmark{j}} & \small{$\left(\frac{g_\mathrm{CALIFA}}{g_\mathrm{SDSS}}\right)$\tablefootmark{k}} & \small{$\left(\frac{r_\mathrm{CALIFA}}{r_\mathrm{SDSS}}\right)$\tablefootmark{l}} & \small{flags(C)\tablefootmark{m}}\\\hline
\endfirsthead
\caption{continued.}\\\hline\hline
\small{ID\tablefootmark{a}} & \small{airmass\tablefootmark{b}} & \small{$\mu_{V,\mathrm{sky}}$\tablefootmark{c}} & \small{$A_V$\tablefootmark{d}} & \small{seeing\tablefootmark{e}} & \small{flags(O)\tablefootmark{f}} & \small{$\delta_\lambda$\tablefootmark{g}} & \small{S/N($R_{50}$)\tablefootmark{h}} & \small{$I_{3\sigma}$\tablefootmark{i}} & \small{flags(R)\tablefootmark{j}} & \small{$\left(\frac{g_\mathrm{CALIFA}}{g_\mathrm{SDSS}}\right)$\tablefootmark{k}} & \small{$\left(\frac{r_\mathrm{CALIFA}}{r_\mathrm{SDSS}}\right)$\tablefootmark{l}} & \small{flags(C)\tablefootmark{m}}\\\hline
\endhead
\hline
\endfoot
\small{001$^*$} & \small{$1.02 \pm 0.01$} & \small{$21.4$} & \small{$0.18$} & \small{...} &\small{000} & \small{$4.68$} & \small{$24.8$} & \small{$24.0~|~0.9$} & \small{00000} & \small{$1.04$} & \small{$1.03$} & \small{000} \\ 
\small{002} & \small{$1.41 \pm 0.05$} & \small{$20.9$} & \small{...} &\small{...} &\small{00--} &\small{$4.63$} & \small{$28.9$} & \small{$23.7~|~1.2$} & \small{00000} & \small{$0.96$} & \small{$0.98$} & \small{000} \\ 
\small{003$^*$} & \small{$1.01 \pm 0.01$} & \small{$20.8$} & \small{$0.23$} & \small{$0.8 \pm 0.1$} & \small{000} & \small{$5.17$} & \small{$12.8$} & \small{$23.7~|~1.2$} & \small{00100} & \small{$0.92$} & \small{$0.97$} & \small{100} \\ 
\small{005} & \small{$1.40 \pm 0.01$} & \small{$21.0$} & \small{...} &\small{...} &\small{00--} &\small{$4.65$} & \small{$24.3$} & \small{$23.6~|~1.3$} & \small{00000} & \small{$0.99$} & \small{$1.01$} & \small{000} \\ 
\small{007$^*$$^\dagger$} & \small{$1.34 \pm 0.05$} & \small{$20.5$} & \small{...} &\small{...} &\small{00--} &\small{$5.21$} & \small{$27.6$} & \small{$23.5~|~1.4$} & \small{10100} & \small{$1.03$} & \small{$1.05$} & \small{000} \\ 
\small{008} & \small{$1.34 \pm 0.07$} & \small{$21.0$} & \small{$0.20$} & \small{...} &\small{000} & \small{$4.63$} & \small{$29.1$} & \small{$23.7~|~1.1$} & \small{00000} & \small{$1.02$} & \small{$0.99$} & \small{000} \\ 
\small{010$^*$} & \small{$1.31 \pm 0.05$} & \small{$21.0$} & \small{$0.22$} & \small{...} &\small{000} & \small{$4.62$} & \small{$18.2$} & \small{$23.7~|~1.2$} & \small{00000} & \small{$1.01$} & \small{$0.98$} & \small{000} \\ 
\small{013} & \small{$1.57 \pm 0.10$} & \small{$20.8$} & \small{$0.19$} & \small{...} &\small{000} & \small{$4.63$} & \small{$31.5$} & \small{$23.6~|~1.3$} & \small{--00--0} & \small{$0.96$} & \small{$1.01$} & \small{001} \\
\end{longtable}
\tablefoot{We describe the meaning of each column including the identifier of each column in the electronic table available on the DR2 web page.
\tablefoottext{a}{IDs marked with an asterisk were already part of the DR1. A dagger indicates cubes that were registered with the old method of the pipeline V1.3c.}
\tablefoottext{b}{Mean airmass (OBS\_AIR\_MEAN) and rms (OBS\_AIR\_RMS) of the observations for the frames used to create the considered datacube.}
\tablefoottext{c}{Average night-sky surface brightness (OBS\_SKY\_MAG) in the $V$ band during the observations in units of mag\,arcsec$^{-2}$.}
\tablefoottext{d}{Average night-sky attenuation (OBS\_EXT\_MEAN) in the V band during the observations in magnitudes.}
\tablefoottext{e}{Average natural seeing (OBS\_SEEING\_MEAN) in the V-band during the observations in arcsec (FWHM).}
\tablefoottext{f}{Observation quality flags, combining the three individual column flags ( FLAG\_OBS\_AM, FLAG\_OBS\_SKYMAG, FLAG\_OBS\_EXT) as described in Sect.~\ref{sect:QC}.}
\tablefoottext{g}{Average spectral resolution (RED\_DISP\_MEAN) in \AA\ (FWHM), measured by fitting the night-sky emission lines with single Gaussian functions.}
\tablefoottext{h}{Average signal-to-noise ratio (CAL\_SNR1HLR) estimated for the full wavelength range at one half light radius from the center.}
\tablefoottext{i}{Average flux  at the 3$\sigma$ continuum detection limit in units of V-band $\mathrm{mag~arcsec}^{-2}$ and in units of $10^{-18}\,\mathrm{erg}\,\mathrm{s}^{-1}\,\mathrm{cm}^{-2}\mathrm{\AA}^{-1}\,\mathrm{arcsec}^{-2}$.}
\tablefoottext{j}{Reduction/instrumental performance quality flags, combining the five individual column flags (FLAG\_RED\_STRAYLIGHT, FLAG\_RED\_DISP, FLAG\_RED\_CDISP, FLAG\_RED\_SKYLINES, FLAG\_RED\_LIMSB) as described in Sect.~\ref{sect:QC}.}
\tablefoottext{k}{Ratio between the SDSS $g$ band flux derived from the datacube and the one derived from the SDSS images for a 30$\arcsec$-diameter aperture (CAL\_QFLUX\_G).}
\tablefoottext{l}{Ratio between the SDSS $r$ band flux derived from the datacube and the one derived from the SDSS images for a 30$\arcsec$-diameter aperture (CAL\_QFLUX\_R).}
\tablefoottext{m}{Quality control flags, combining the three individual column flags (FLAG\_CAL\_SPECPHOTO, FLAG\_CAL\_WL, FLAG\_CAL\_IMA) as described in Sect.~\ref{sect:QC}.}
}
\end{longtab}

\begin{longtab}
\begin{longtable}{lcccccccccc}
\caption{CALIFA DR2 quality control parameters for the V1200 data }\\\hline\hline
\label{tab:QC_par_V1200}
\small{ID\tablefootmark{a}} & \small{airmass\tablefootmark{b}} & \small{$\mu_{V,\mathrm{sky}}$\tablefootmark{c}} & \small{$A_V$\tablefootmark{d}} & \small{seeing\tablefootmark{e}} & \small{flags(O)\tablefootmark{f}} & \small{$\delta_\lambda$\tablefootmark{g}} & \small{S/N($R_{50}$)\tablefootmark{h}} & \small{$I_{3\sigma}$\tablefootmark{i}} & \small{flags(R)\tablefootmark{j}} & \small{flags(C)\tablefootmark{k}}\\\hline
\endfirsthead
\caption{continued.}\\\hline\hline
\small{ID\tablefootmark{a}} & \small{airmass\tablefootmark{b}} & \small{$\mu_{V,\mathrm{sky}}$\tablefootmark{c}} & \small{$A_V$\tablefootmark{d}} & \small{seeing\tablefootmark{e}} & \small{flags(O)\tablefootmark{f}} & \small{$\delta_\lambda$\tablefootmark{g}} & \small{S/N($R_{50}$)\tablefootmark{h}} & \small{$I_{3\sigma}$\tablefootmark{i}} & \small{flags(R)\tablefootmark{j}} & \small{flags(C)\tablefootmark{k}}\\\hline
\endhead
\hline
\endfoot
\small{001$^*$} & \small{$1.01 \pm 0.02$} & \small{...} &\small{$0.26$} & \small{...} &\small{0--0} & \small{$1.90$} & \small{$ 9.0$} & \small{$23.2~|~3.3$} & \small{00000} & \small{000} \\ 
\small{002} & \small{$1.32 \pm 0.04$} & \small{$21.8$} & \small{$0.13$} & \small{...} &\small{000} & \small{$1.94$} & \small{$12.4$} & \small{$23.3~|~3.0$} & \small{00000} & \small{000} \\ 
\small{003$^*$} & \small{$1.08 \pm 0.06$} & \small{$22.0$} & \small{$0.17$} & \small{...} &\small{000} & \small{$1.96$} & \small{$ 6.5$} & \small{$23.3~|~2.9$} & \small{11000} & \small{000} \\ 
\small{005} & \small{$1.44 \pm 0.05$} & \small{$21.8$} & \small{$0.16$} & \small{...} &\small{000} & \small{$1.94$} & \small{$13.2$} & \small{$23.4~|~2.8$} & \small{--00--0} & \small{100} \\ 
\small{007$^*$$^\dagger$} & \small{$1.18 \pm 0.02$} & \small{$21.7$} & \small{$0.14$} & \small{...} &\small{000} & \small{$1.96$} & \small{$10.2$} & \small{$22.8~|~4.9$} & \small{00000} & \small{001} \\ 
\small{008} & \small{$1.04 \pm 0.03$} & \small{$21.9$} & \small{$0.43$} & \small{...} &\small{001} & \small{$1.90$} & \small{$10.2$} & \small{$23.2~|~3.3$} & \small{00010} & \small{000} \\ 
\small{010$^*$} & \small{$1.29 \pm 0.09$} & \small{$22.2$} & \small{...} &\small{...} &\small{00--} &\small{$1.92$} & \small{$ 6.1$} & \small{$23.0~|~4.1$} & \small{00000} & \small{001} \\ 
\small{013} & \small{$1.53 \pm 0.04$} & \small{$21.8$} & \small{$0.14$} & \small{...} &\small{000} & \small{$1.94$} & \small{$16.0$} & \small{$23.4~|~2.9$} & \small{00010} & \small{100} \\
\end{longtable}
\tablefoot{We describe the meaning of each column including the identifier of each column in the electronic table available on the DR2 web page.
\tablefoottext{a}{IDs marked with an asterisk were already part of the DR1. A dagger indicates cubes that were registered with the old method of the pipeline V1.3c.}
\tablefoottext{b}{Mean airmass (OBS\_AIR\_MEAN) and rms (OBS\_AIR\_RMS) of the observations for the frames used to create the considered datacube.}
\tablefoottext{c}{Average night-sky surface brightness (OBS\_SKY\_MAG) in the $V$ band during the observations in units of mag\,arcsec$^{-2}$.}
\tablefoottext{d}{Average night-sky attenuation (OBS\_EXT\_MEAN) in the V band during the observations in magnitudes.}
\tablefoottext{e}{Average natural seeing (OBS\_SEEING\_MEAN) in the V-band during the observations in arcsec (FWHM).}
\tablefoottext{f}{Observation quality flags, combining the three individual column flags ( FLAG\_OBS\_AM, FLAG\_OBS\_SKYMAG, FLAG\_OBS\_EXT) as described in Sect.~\ref{sect:QC}.}
\tablefoottext{g}{Average spectral resolution (RED\_DISP\_MEAN) in \AA\ (FWHM), measured by fitting the night-sky emission lines with single Gaussian functions.}
\tablefoottext{h}{Average signal-to-noise ratio (CAL\_SNR1HLR) estimated for the full wavelength range at one half light radius from the center.}
\tablefoottext{i}{Average flux  at the 3$\sigma$ continuum detection limit in units of B-band $\mathrm{mag~arcsec}^{-2}$ and in units of $10^{-18}\,\mathrm{erg}\,\mathrm{s}^{-1}\,\mathrm{cm}^{-2}\mathrm{\AA}^{-1}\,\mathrm{arcsec}^{-2}$.}
\tablefoottext{j}{Reduction/instrumental performance quality flags, combining the five individual column flags (FLAG\_RED\_STRAYLIGHT, FLAG\_RED\_DISP, FLAG\_RED\_CDISP, FLAG\_RED\_SKYLINES, FLAG\_RED\_LIMSB) as described in Sect.~\ref{sect:QC}.}
\tablefoottext{k}{Quality control \textsc{cal} flags, combining the three individual column flags (FLAG\_CAL\_SPECPHOTO, FLAG\_CAL\_WL, FLAG\_CAL\_IMA) as described in Sect.~\ref{sect:QC}.}
}
\end{longtab}

\subsection{Quality of the instrumental/data reduction performance (\textsc{red})}\label{subsect:QC_red} 

The quality of the instrumental and data reduction performance is assessed via a series of four quantities measured 
on the reduced data {\em before} combining them into the final datacube: \textsc{straylight}, spectral 
\textsc{dispersion}, cross dispersion \textsc{cdisp}, and the residuals from the subtraction of bright skylines 
(namely, the 5577\AA~O$_2$~line in the V500 setup and the 4358\AA~Hg\textsc{i}~in the V1200 setup). In addition 
we consider the limiting surface brightness corresponding to a 3-$\sigma$ detection measured on the final cube.

The so-called straylight is an additional source of illumination internal to the instrument, possibly as 
a distributed scattered light component. Straylight, if not subtracted properly, introduces systematic errors 
and thus limits the final sensitivity and accuracy of the data reduction\footnote{For a detailed description on 
the straylight subtraction, see Appendix A.3 of \citetalias{Husemann:2013}.}. High mean 
level of straylight in a frame (\textsc{meanstraylight}), as well as high maximum values (\textsc{maxstraylight}) 
and large rms (\textsc{rmsstraylight}), are indication of poor performance. Levels above the thresholds provided 
in Tables \ref{tab:QCflags_def_V500} and \ref{tab:QCflags_def_V1200} in at least one of the exposures 
(\textsc{\_max} suffix) raise a \textsc{warning} or a \textsc{bad} \textsc{flag\_red\_straylight} flag.

The spectral dispersion and cross dispersion are measured on individual fiber spectra as the FWHM of skylines and 
the FWHM of the spectral trace, respectively. Thresholds are set on the mean values to ensure that the typical 
parameters do not depart too much from the nominal target specifications, and on the maximum and rms in order to 
check for anomalies in the data. Any failure to comply within the thresholds reported in Tables 
\ref{tab:QCflags_def_V500} and \ref{tab:QCflags_def_V1200} raises a \textsc{flag\_red\_disp} 
or \textsc{flag\_red\_cdisp}.

In order to assess the performance of the sky subtraction we consider the minimum and the maximum over all 
pointings of the average (over all fibers) flux residual of a bright skyline within an individual pointing 
(\textsc{red\_res4358\_min} and \textsc{red\_res4358\_max}, and \textsc{red\_res5577\_min} and 
\textsc{red\_res5577\_max} for the V1200 and the V500 setup respectively). We also consider the maximum 
over all pointings of the rms residuals (over all fibers in an individual pointing), 
\textsc{red\_rmsres4358\_max} and \textsc{red\_rmsres5577\_max}. Too negative or too positive average 
residuals are indication of systematic bias in the sky subtraction. Too large rms can be regarded as 
symptom of localized failures or noisy data. In these cases the \textsc{flag\_red\_skylines} is set.

Finally, the 3$\sigma$ continuum flux density detection limit per interpolated 1 arcsec$^2$-spaxel for 
the faintest regions is used to identify cubes whose depth does not fulfil the survey requirements and 
this is reflected in the \textsc{flag\_red\_limsb} flag. More about the depth of the final datacubes is 
discussed in Sec.~\ref{sect:depth}.

\begin{figure}
\resizebox{\hsize}{!}{\includegraphics{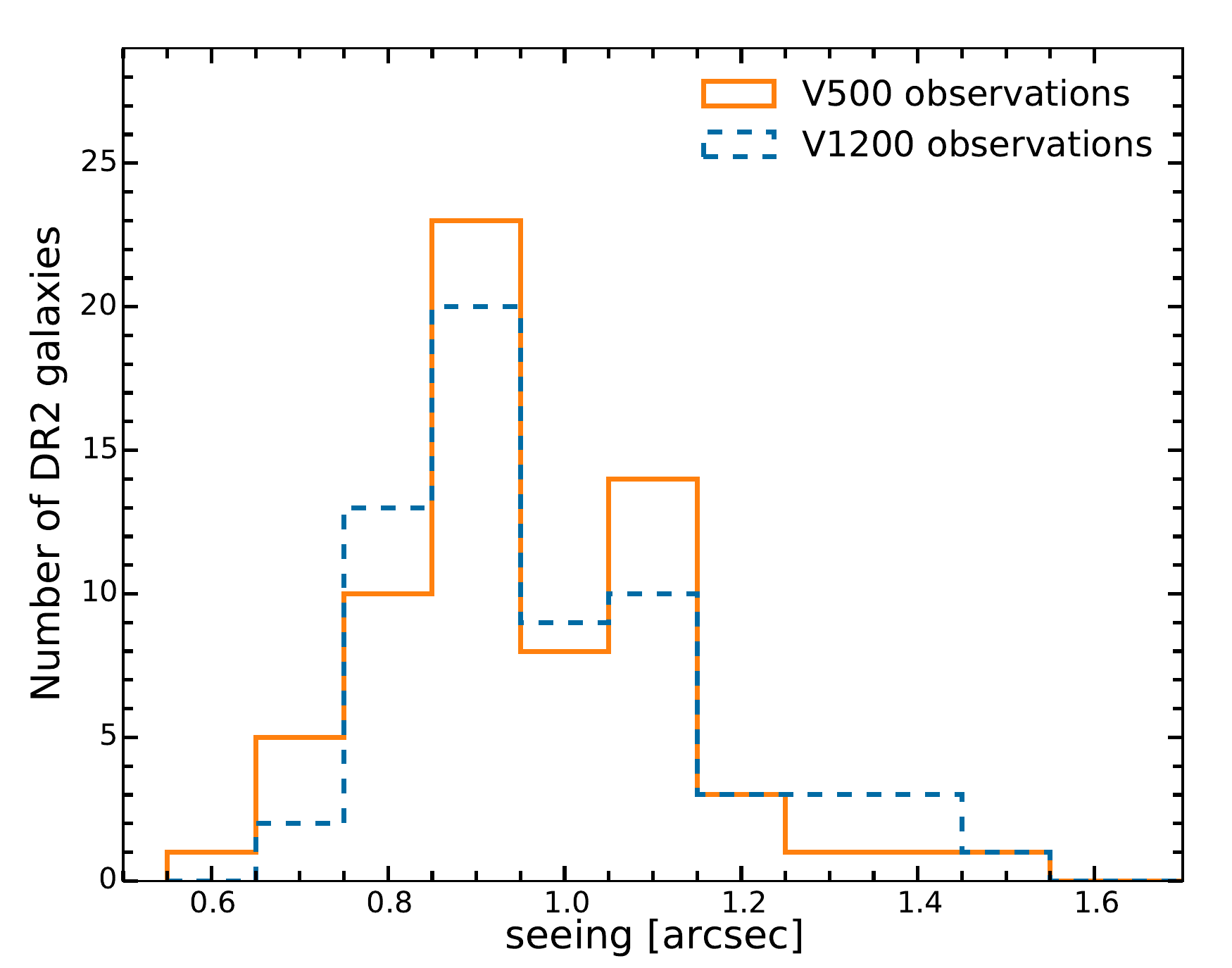}}
\caption{Distribution of the seeing during the CALIFA observations as measured
  by the automatic Differential Image Motion Monitor
  \citep[DIMM,][]{Aceituno:2004}.}
  \label{fig:DR2_seeing}
\end{figure}

\subsection{Quality of the calibrated data products (\textsc{cal})}\label{subsect:QC_cal}

The quality of the calibrated data products is determined by checks on the global spectrophotometry, on 
the stability of the wavelength calibration across the spectral range, and on the quality of the 
resulting 2D flux distribution (synthetic image) and its ability to match the SDSS broad-band imaging.

\begin{figure}
\resizebox{\hsize}{!}{\includegraphics{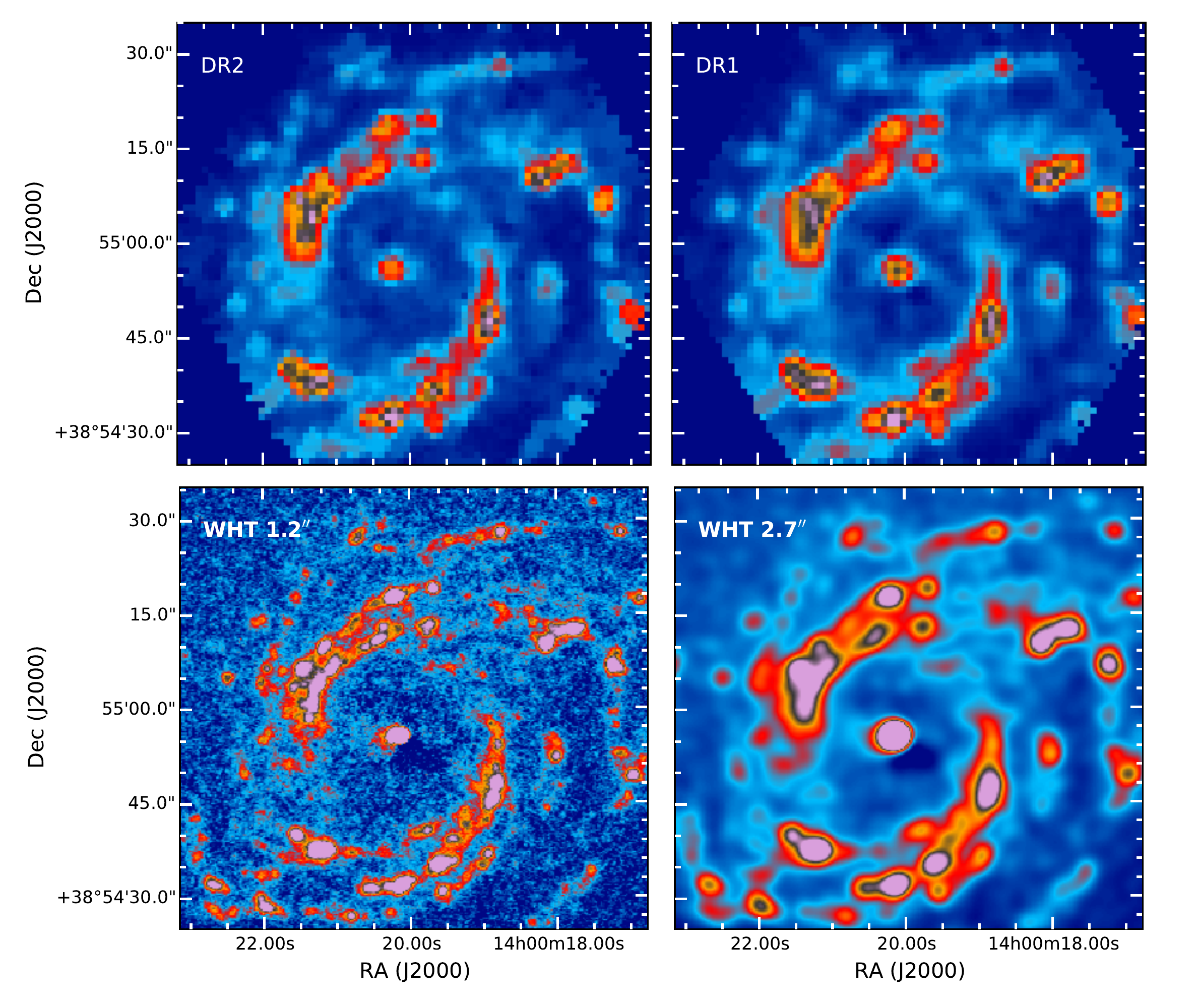}}
\caption{DR2 spatial resolution comparison for NGC 5406 (ID 684). The upper left panel shows the DR2 
	image of the H$\alpha$ map and the upper right the DR1 one. The lower row are H$\alpha$ images 
	taken with the 4.2m William Herschel Telescope (Roque de los Muchachos Observatory, La Palma, Spain), 
	using the AUXCAM detector (S\'anchez-Menguiano et al., in prep.). The image, with an original 
	resolution of 1.2\arcsec\ (bottom left), has been degraded to a resolution of 2.7\arcsec\ 
        (bottom right) and the FoV has been reduced to match exactly the same WCS coordinates as CALIFA.}
  \label{fig:DR2_spatial}
\end{figure}

The quality of global spectrophotometric calibration is assessed by comparing the photometric fluxes 
derived from spectra integrated within 30\arcsec -radius apertures with the corresponding fluxes 
derived from SDSS imaging, as explained below in Sec.~\ref{sect:specphot_cal}. For the V500 setup, 
in particular, it is possible to derive the flux ratio between SDSS and CALIFA in $g$ and $r$-band 
(\textsc{cal\_qflux\_g} and \textsc{cal\_qflux\_r}, respectively, averaged over all pointings 
for a given galaxy): values of these ratios departing from 1 by more than the tolerances listed 
in Table \ref{tab:QCflags_def_V500} are flagged. Large rms variations of these values over the 
three V500 pointings (\textsc{cal\_qflux\_rms}, which combines $g$ and $r$ bands) are also considered 
symptoms of poor quality. In addition to these quantitative parameters, we visually check 
that the spectral energy distribution (SED) measured via SDSS photometry matches the CALIFA integrated 
spectrum. For this check we also consider the $u$ and the $i$ band data-points: although the CALIFA 
spectra do not cover the full extent of these pass-bands, they prove helpful in judging the matching 
of spectral shapes. Five members of the collaboration have performed these checks independently and 
assigned flags \textsc{ok-warning-bad}: the second-to-worst classification is retained. This flag is 
then combined with the flags based on the quantitative flux ratios to create the final 
\textsc{flag\_cal\_specphoto} flag.

In order to check the stability of the wavelength calibration over the full spectral range we 
performed the same measurements presented in Sec. 5.3 of \citetalias{Husemann:2013}: for each 
galaxy and setup, the spectra within 5\arcsec\ of the center of the galaxy are integrated and the 
systemic velocity is estimated first for the full spectrum and then for 3 (4) independent spectral 
ranges in V1200 (V500); the rms of these values with respect to the systemic velocity from the full 
spectrum (\textsc{cal\_rmsvelmean}) is an estimate of the stability of the wavelength calibration 
across the wavelength range and is used to set the corresponding quality flag \textsc{flag\_cal\_wl}. 
In $>97.5$\% of the cases we obtain \textsc{cal\_rmsvelmean} well below 2 km sec$^{-1}$ for the V1200 
and 3 km sec$^{-1}$ for the V500 grating.

Finally, the quality flag on the 2D flux distribution and plane-of-sky registration, \textsc{flag\_cal\_ima}, 
is defined by combining the information on the goodness of matching between SDSS images and synthetic images 
from the CALIFA datacube and a series of visual checks. The former piece of information is provided by the 
chi-squared of the registration procedure (see Sect.~\ref{sect:pipeline}). The visual checks include: a check 
on possible artefacts in the synthetic broad band image from the final CALIFA cubes (e.g. mismatched 
features, elongated PSF); a comparison of the CALIFA fiber footprints of each pointing with the registered 
SDSS image, looking for apparent mismatched and miscomputed spatial offsets; a check of the chi-squared 
surface plot displaying the dependence of the registration procedure (see below) on the x and y spatial 
offsets, whereby irregular chi-square surfaces and lacks of clear minimum likely imply the impossibility 
of an accurate registration. Out of five independent classifiers we chose the median value of the 
attributed flags and combine it with the flag corresponding to the chi-squared measurements. We note 
that a small number of objects already released as part of DR1 do not reach the imaging quality standards 
using the registration procedure adopted in the pipeline V1.5 (see Sect.~\ref{sect:pipeline}), which uses 
cross-correlation with SDSS images: in these cases we revert to the old registration scheme adopted for 
DR1 (pipeline V1.3c) and mark the objects with a dagger in Tables \ref{tab:QC_par_V500} and 
\ref{tab:QC_par_V1200}.

\begin{figure}
\resizebox{\hsize}{!}{\includegraphics{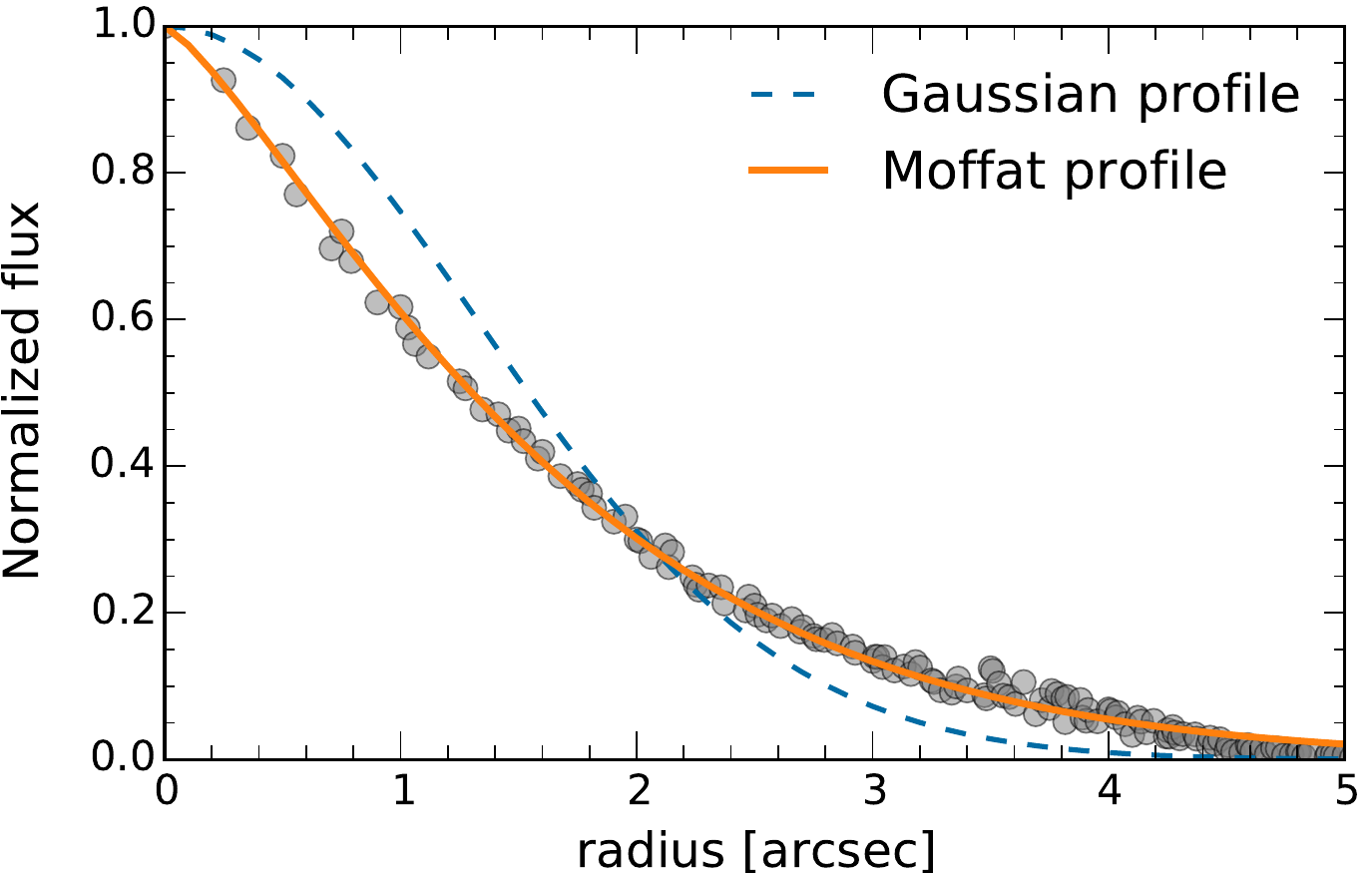}}
\caption{Profile fit to a foreground star close to the center in the data 
   cube of the galaxy NGC 2916. 
   When the PSF is good, a Moffat function fits the data better than a Gaussian.}
  \label{fig:DR2_PSF_GM}
\end{figure}

\begin{figure}
\resizebox{\hsize}{!}{\includegraphics{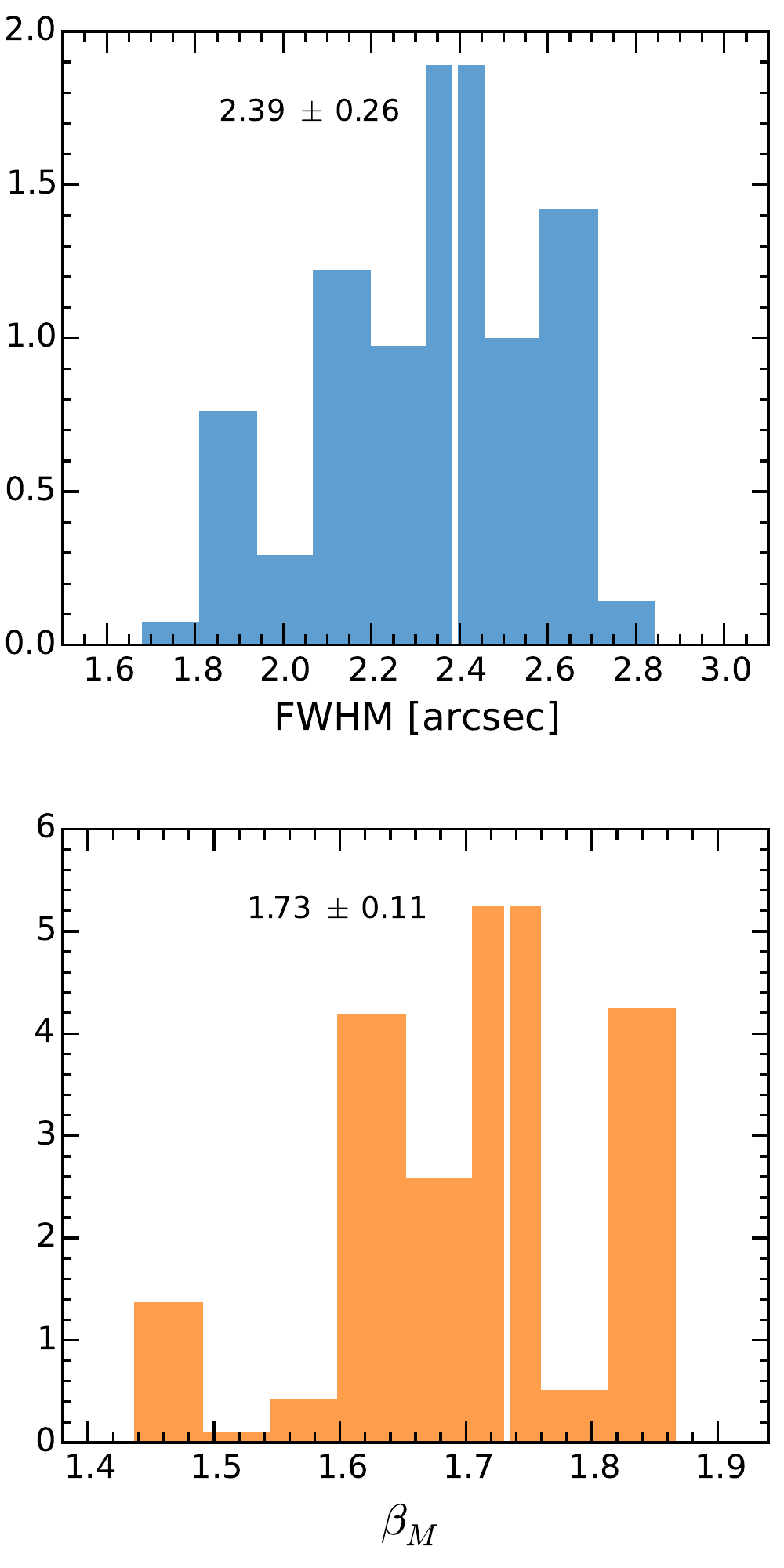}}
\caption{Normalized Distribution of PSF FWHM (\emph{top}) and $\beta_M$ (\emph{bottom}) parameters of a 2-d 
    Moffat profile fitted to 45 calibration stars, weighted by the likelihood of the fit. The mean 
    value of the distributions are marked with a white solid line.}
  \label{fig:DR2_PSF_stars}
\end{figure}

\subsection{Astrometric accuracy and spatial resolution}\label{sect:astrometry}

\subsubsection{Astrometric registration accuracy}\label{sect:astrometry}

Pipeline V1.5 implements a new method (see Sect.~\ref{sect:pipeline}) to register the absolute 
astrometry of the datacube coordinate system to the International Coordinate Reference 
System (ICRS). The previous pipeline, V1.3c, used tabulated coordinates of the galaxy $V$ band 
photometric center that were assigned to the barycenter measured in the reconstructed image from 
the datacubes (just one point, instead of the global match applied in V1.5).

In order to check the accuracy of the new astrometric registration for V500 and V1200 datacubes, we 
performed independent tests using SDSS $r$ and $g$-band images (DR10) for each galaxy. 
Synthetic $r$ and $g$-band PPaK images were computed using the V1.5 reduced data. The coordinates of 
the peak centroid PPAK images are used as an approximate galactic center, and the corresponding 
peak was measured in the SDSS images. The offsets between the SDSS and CALIFA are less 
than 3\arcsec\ (rms $\sim$ 1\arcsec) for the majority of the DR2 sample.  
Large offsets are mostly due either to edge-on galaxies, centers of the 
galaxies not well defined due to dust lanes, irregular morphology or bright field star(s) near the 
center of the galaxy. Objects with offsets larger than 3\arcsec\ measured in V500 setup are: IC1652, 
NGC0444, UGC00809, UGC00841, NGC0477, IC1683, NGC0499, NGC0496, NGC0528, UGC01938, NGC1056, NGC3991, 
MCG-01-01-012 and NGC7800. For the V1200 setup: IC1528, IC1652, NGC0444, UGC00809, UGC00841, NGC0477, 
NGC0499, NGC0496, NGC0528, UGC02222, NGC3991, UGC11792, MCG-01-01-012 and NGC7800.

\subsubsection{Seeing and spatial resolution}\label{sect:spatial_resolution}

In order to cover the complete FoV of the central bundle and to increase the final resolution of the 
CALIFA datacubes (PPak fibers have a diameter of 2.7\arcsec), a dithering scheme with three pointings 
has been adopted, as described in \citetalias{Sanchez:2012a}. In imaging, in addition to the telescope 
aperture, instrumental and atmospheric seeing determine the final spatial resolution. This has to be 
added to the IFU particular characteristics.

The average atmospheric seeing conditions along the time of the observation of the CALIFA data
were derived from the measurements acquired by the DIMM \citep[DIMM,][]{Aceituno:2004}, 
which operates fully automatically at the Calar Alto observatory during the night. 
DIMM has different operational constraints (humidity lower 
than 80\% and wind speed less than $12\,\mathrm{m}\,\mathrm{s}^{-1}$) than the 3.5m 
telescope, thus seeing information is not available for every CALIFA observation. Thus,  
there can be some DIMM seeing values missing from Tables~\ref{tab:QC_par_V500} and \ref{tab:QC_par_V1200}, 
but the overall seeing distribution is not expected to be very different. Figure~\ref{fig:DR2_seeing}
shows the DIMM seeing distribution for the DR2 sample, which has a median value of 
$0\farcs9$ FWHM (the distribution is very similar to the DR1 sample), and therefore atmospheric 
seeing is not a limiting factor in the spatial resolution of the CALIFA cubes. 

Another improvement of the pipeline, as discussed \ref{sect:pipeline}, is the spatial resolution. 
Fig.~\ref{fig:DR2_spatial} shows H$\alpha$ maps (obtained using FIT3D on the CALIFA datacubes) for 
NGC 5406 (ID 684) for DR1, DR2 and one image taken with the William Herschel Telescope (WHT) using 
a narrowband filter. The last image has been also degraded to the DR2 nominal resolution for the 
sake of comparison. This improvement impacts directly, for example, onto the detection rate of 
\ion{H}{ii} regions. Using {\sc HIIexplorer} \citep{Sanchez:2012b} on the V1.3c 
datacubes of the 200 galaxies, a total of 5878 are recovered, while this number rises to 
7646 \ion{H}{ii} regions for the DR2 galaxies using pipeline V1.5, which represents an increase 
of $\sim$ 30\%.

\begin{figure}
\resizebox{\hsize}{!}{\includegraphics{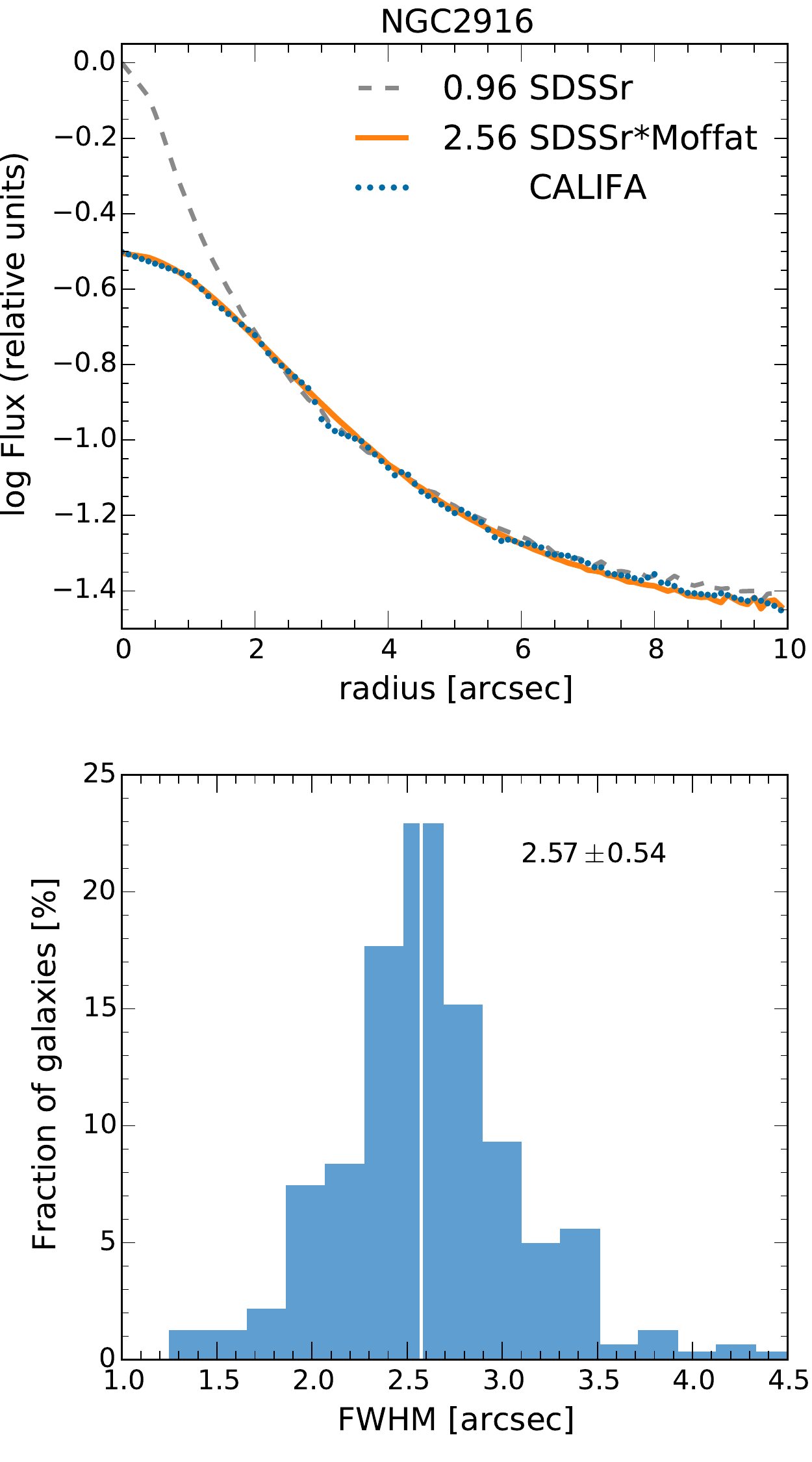}}
\caption{\emph{Top panel:} The dashed (grey) line is the flux profile of the galaxy NGC 2916 from the 
   original SDSS $r$-band image, with a PSF FWHM = 0\farcs96 (as obtained from the SDSS Skyserver), 
   the dotted (blue) line is the flux profile from the CALIFA datacube of NGC 2916, and the solid  
   (orange) line is the profile from the convolution of the SDSS $r$ image with a 2D Moffat kernel of 
   width 2\farcs37, resulting in a final PSF FWHM = 2\farcs56.
   \emph{Bottom panel:} Distribution of FWHM obtained for all galaxies when convolving 
   SDSS $r$ images with a Moffat kernel, with values around $2.57\pm0.54$ arcsec (white line).}
  \label{fig:DR2_PSF}
\end{figure}

We follow two different approaches to measure the PSF in the datacubes. In those cases where a 
bright foreground star is present not far from the center, a fit is performed to the radial distribution 
of the star light profile. When the PSF is good, a Moffat function yields a better fit than the Gaussian, 
as shown in Fig. \ref{fig:DR2_PSF_GM}. This method can only be applied to a few galaxies, usually 
with stars being far from the center of the images, by construction of the survey. 

Since January 2012 standard stars were observed using the same dithering pattern adopted for the science 
observations for the V500 setup. We observed a total of 107 nights in this period. Only 70\% of the 
nights had weather conditions good enough to acquire a calibration star and 2/3 were observed 
adopting the dithering scheme, yielding a total of 45 datacubes. We reduced the data using the same 
procedure described before for the science objects. As the calibration stars have a very high S/N, 
the PSF can be measured very precisely. From the datacubes we simulate a SDSS $g$-band image for 
each of these stars. For each of these images we fit a 2-d Moffat profile using the software 
{\sc IMFIT} \citep{Erwin:2014}\footnote{\url{http://www.mpe.mpg.de/~erwin/code/imfit/}}.
Figure~\ref{fig:DR2_PSF_stars} shows the normalized distributions of FWHM and $\beta_M$ parameters 
of the Moffat profile, weighted by the likelihood of the fit. We obtain a mean value of the 
FWHM = 2.39 $\pm$ 0.26 arcsec, with $\beta_M$ = 1.73 $\pm$ 0.11. We also measure an ellipticity 
(1 - b/a, with a and b being the semi-major and semi-minor axes, respectively) of 0.08 $\pm$ 0.06. 
Given the uncertainties, this means the PSF can be considered effectively axisymmetric. The 
uncertainties in these measurements correspond to 1-$\sigma$ of the distributions.

\begin{figure*}
 \includegraphics[width=0.5\textwidth]{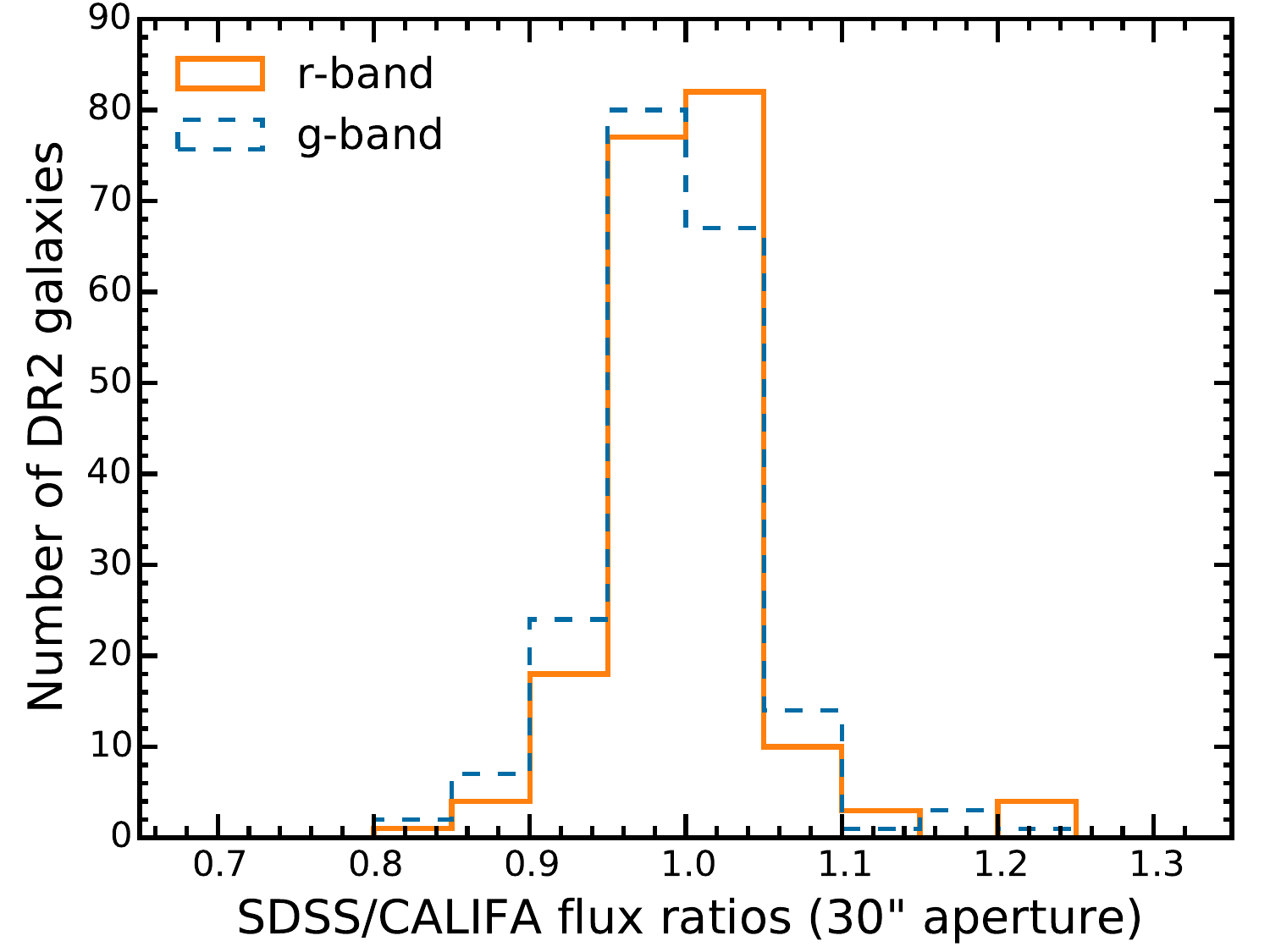}
 \includegraphics[width=0.5\textwidth]{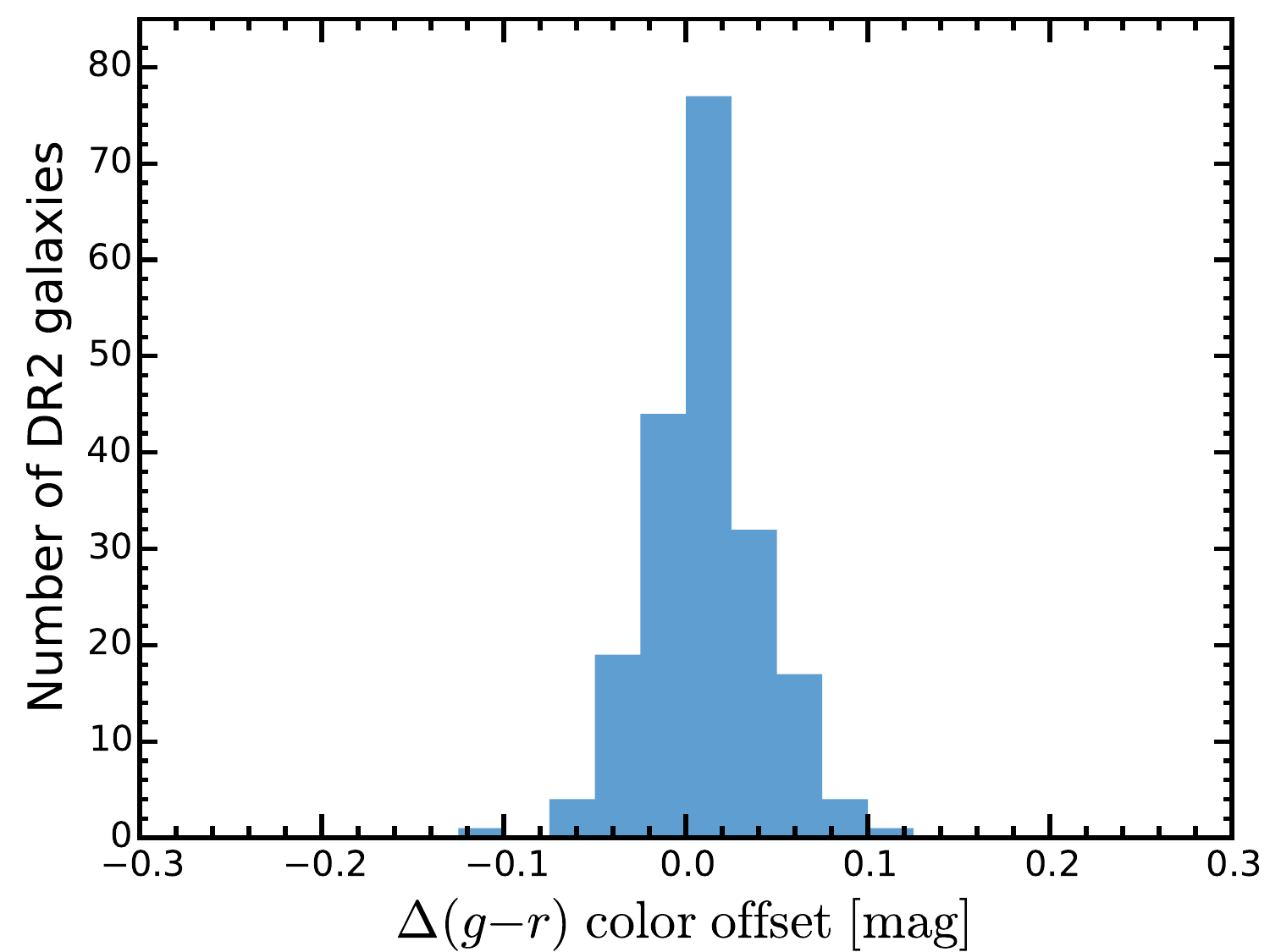}

 \caption{\emph{Left panel:} Distribution of the 30\arcsec\ aperture
   photometry scale factor between the SDSS DR7 images and re-calibrated
   CALIFA data. We compare the photometry only for the $g$ and $r$ bands,
   which are both entirely covered by the V500 wavelength range. \emph{Right
     panel:} Distribution of the corresponding color offset between the SDSS
   DR7 images and the synthetic CALIFA broad-band images.  }
  \label{fig:DR2_scale_check}
\end{figure*}

The second method is based on the fact that the nuclei of galaxies 
have a relatively steep luminosity profile, and we take advantage of the good quality of the SDSS 
images. For each galaxy, the SDSS $r$-band image is convolved with a 2D function (Moffat or Gaussian) 
and we compute the residuals after subtracting the resulting flux radial distribution from the flux 
radial distribution of the datacube of the same galaxy. A range of FWHM for the convolution 
function is used and we choose the one that minimizes the residuals. An example for the galaxy 
NGC 2916 is shown in the {top panel} of Fig.~\ref{fig:DR2_PSF} for the case of convolving with 
a Moffat kernel. The dashed (grey) line is the flux profile of the original SDSS $r$-band image, 
with a PSF FWHM = 0\farcs96 (as obtained from the SDSS Skyserver), the dotted (blue) line is the 
flux profile from the CALIFA datacube, and the continuous (orange) line is the convolution of the 
grey profile with a 2D Moffat kernel of width 2\farcs37, resulting in a final PSF FWHM = 2\farcs56. 
The \emph{bottom panel} of Fig.~\ref{fig:DR2_PSF} shows the distribution of FWHM obtained for all galaxies 
for the case of convolving with a Moffat kernel, with values around $2.57\pm0.54$ arcsec (white line).

Both methods, even though being from completely different natures, agree with each other in the value 
of FWHM. Thus, we conclude that a Moffat profile with FWHM = 2.5\arcsec\ and $\beta_M$ = 1.7 is a 
good description of the PSF of the datacubes in this data release.


\subsection{Spectrophotometric accuracy}\label{sect:specphot_cal}

As mentioned in Sect.~\ref{sect:pipeline} the new registration scheme of the pipeline uses 
SDSS $r$-band for the V500 setup and $g$-band for the V1200, and field calibration images 
of the SDSS DR7. Each V500 datacube is rescaled in the absolute flux level to match the 
SDSS DR7 broad-band photometry using the photometric scale factor at the best matching position 
for each pointing. On the other hand, the V1200 data is matched to the V500 data 
(\citetalias{Sanchez:2012a}). This procedure, together with the new re-calibrated sensitivity 
curve (see Sect.~\ref{sect:pipeline} and Husemann et al., in preparation), improves the 
spectrophotometric calibration over DR1. This is clearly shown in Fig.~\ref{fig:DR2_scale_check}. 
As part of the CALIFA pipeline V1.5, a 30\arcsec\ diameter photometric aperture in $r$ and $g$ 
is measured both in the SDSS DR7 images and the equivalent synthetic CALIFA broad-band images. 
The mean SDSS/CALIFA $g$ and $r$ band ratios in DR2 are 1.00 $\pm$ 0.05 and 0.99 $\pm$ 0.06, 
respectively. In the \emph{right panel} of Fig.~\ref{fig:DR2_scale_check} the distribution 
in $\Delta(g-r)$ color difference between the SDSS and CALIFA data shows that the 
spectrophotometric accuracy across the wavelength range is better that 3\%, with a median 
value of  0.01 $\pm$ 0.03.

Spectral fitting methods can be used to perform useful tests of the data and their calibration, and this has 
been done before in CALIFA. \citetalias{Husemann:2013} used {\sc starlight} fits to evaluate the accuracy of 
the error estimates in DR1 datacubes, while \citet{CidFernandes:2014} used such fits to map systematic features 
in the spectral residuals which may indicate calibration issues.

\begin{figure*}
 \resizebox{0.99\textwidth}{!}{\includegraphics{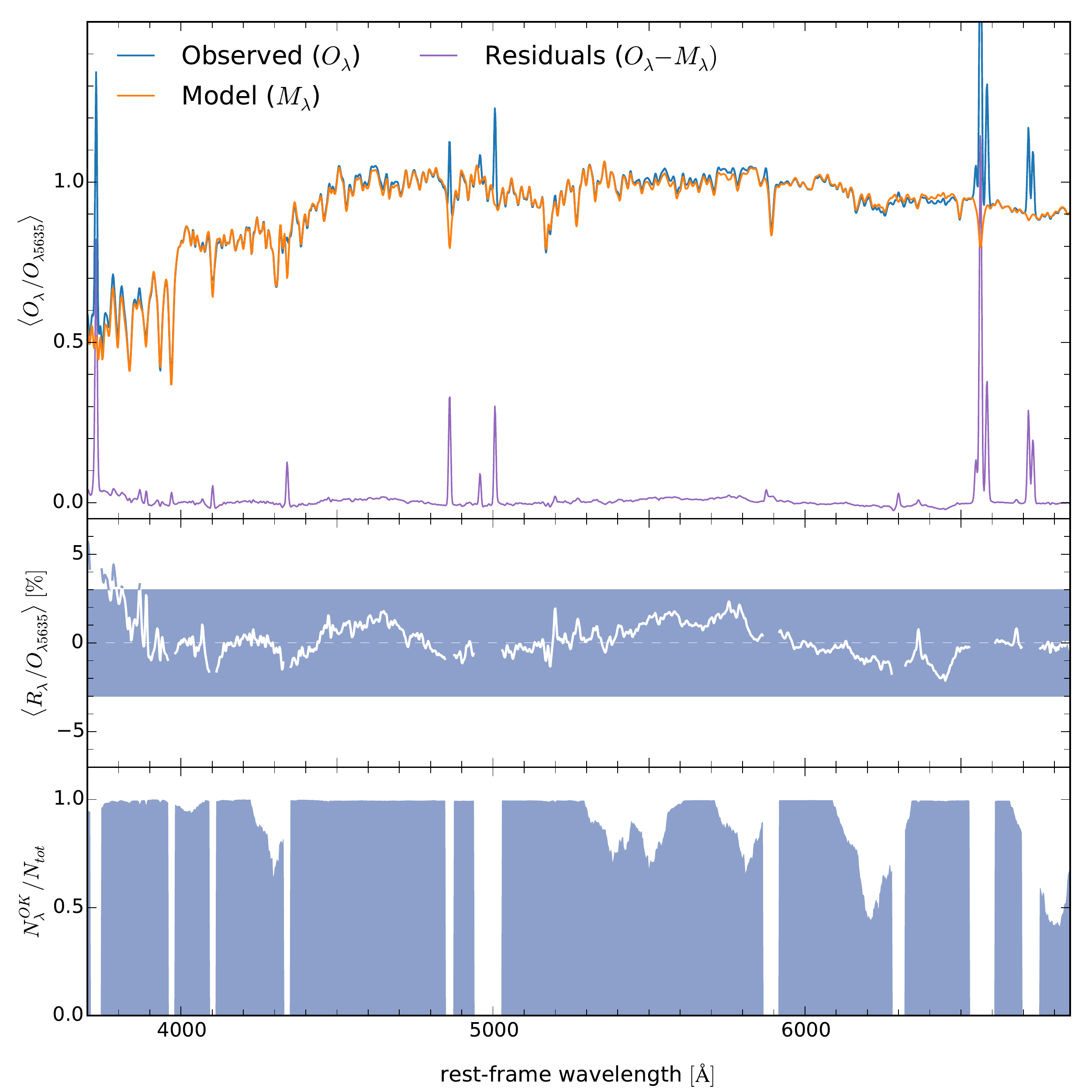}}
 \caption{Statistics of the spectral residuals \citep[compare to Fig.~13 of][]{CidFernandes:2014}. {\em Top:} 
    The mean normalized spectrum of 170670 bins from 200 galaxies. The mean {\sc starlight} fit is overplotted 
    in orange, while the mean residual is plotted at the bottom of the panel (purple). 
    {\em Middle:} Zoom of the residual spectrum, with emission lines removed for clarity. The shaded rectangle 
    encompasses the $\pm$ 3\% area.
    {\em Bottom:} Fraction of the bins contributing to the statistics at each $\lambda$.}
  \label{fig:DR2_residuals}
\end{figure*}

We have repeated the same experiments for the DR2 datacubes. Results are shown in Fig.~\ref{fig:DR2_residuals}. 
The top panel shows in blue the mean spectrum of 170670 Voronoi bins of the 200 galaxies in DR2\footnote{The 
spatial binning is used to guarantee a minimum $S/N$ of 20 in the continuum at $\sim 5635$ \AA. In practice, 88\% 
of the Voronoi bins actually comprise a single spaxel.}. The average is done after normalizing each spectrum 
by its median flux in the $5635 \pm 45$ \AA\ window. The mean synthetic spectrum (overplotted orange line) as well 
as the mean residual (at the bottom of the upper panel, purple line) are also plotted. The middle panel zooms in 
on the residual spectrum, which now excludes emission lines and bad-pixels masked away in the fitting process. 
Finally, the bottom panel shows what fraction of all spectra was used in the statistics at each $\lambda$.

The layout of Fig.~\ref{fig:DR2_residuals} is identical to figure~13 of \citet{CidFernandes:2014}, to which it 
should be compared\footnote{Fig.~13 in \citet{CidFernandes:2014} is in fact busier than our 
Fig.~\ref{fig:DR2_residuals}, as it shows results obtained with three different spectral bases. Here we adopt the 
same base described in \citet{GonzalezDelgado:2014a}, which is very similar to base {\em GM} in 
\citet{CidFernandes:2014}.}. Focusing on the middle panel, one sees that from $\sim 5000$ \AA\ to the red the 
residuals are very similar, including the humps around 5800 \AA, associated to imperfect removal of telluric 
features. Towards the blue however, the new reduction pipeline leads to smaller residuals. For 
instance, the broad feature around H$\beta$ seen with V1.3c spectra is essentially gone with the new reduction. 
A systematic excess blueness persists for $\lambda < 3900$ \AA, but overall the improvement is clear.

\begin{figure*}
 \resizebox{0.99\textwidth}{!}{\includegraphics{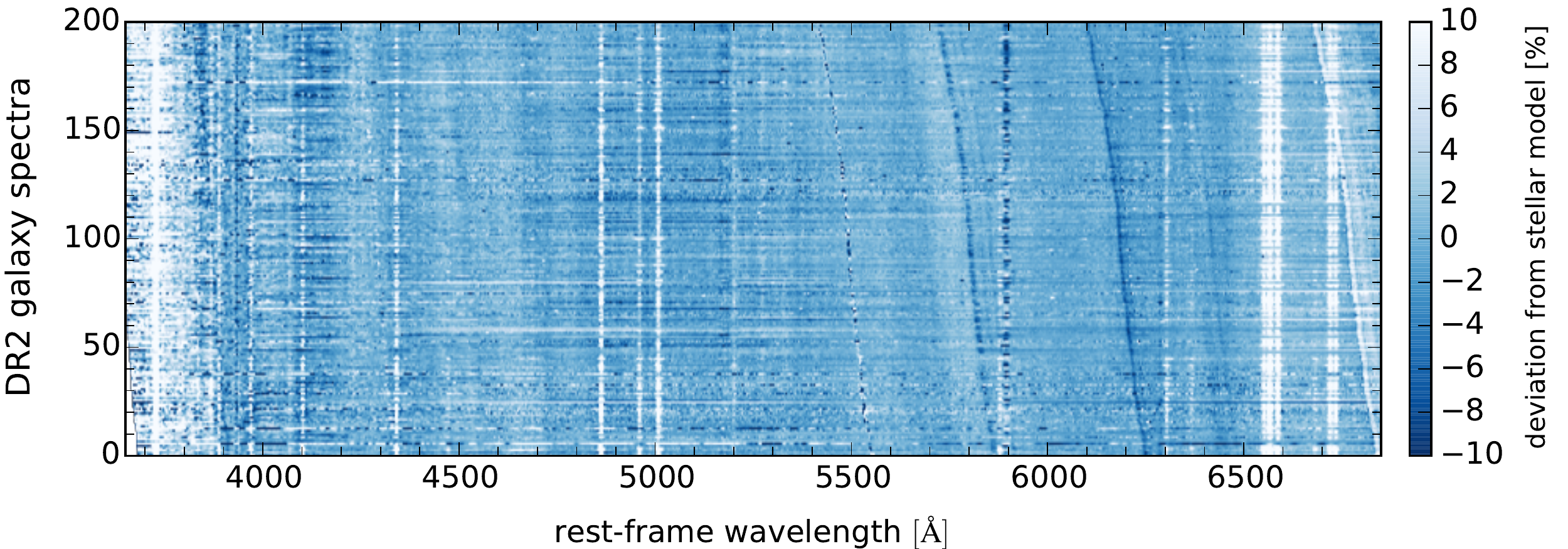}}
 \caption{Relative spectral deviations ---$(O_\lambda - M_\lambda) / O_\lambda$, where $O$ and $M$ denote the observed 
   and the model spectra, respectively--- for the nuclear regions of all DR2 galaxies, vertically sorted by redshift. 
   Unlike in Fig.~\ref{fig:DR2_residuals}, emission lines and bad-pixels are not masked in this plot. Systematic 
   deviations from the {\sc starlight} model appear as vertical stripes (rest-frame mismatches, e.g. imperfect 
   stellar model or emission lines), while slanted stripes trace observed-frame mismatches (e.g. imperfect sky model). 
   Compare to Fig.~16 of \citetalias{Husemann:2013}.}
 \label{fig:DR2_stacked_residuals}
\end{figure*}

Residuals for the 200 DR2 nuclear spectra are shown in Fig.~\ref{fig:DR2_stacked_residuals}, where galaxies are 
sorted by redshift and piled up. This visualization facilitates the identification of telluric features, which 
appear as slanted lines in the image. Comparison with an identical plot in \citetalias{Husemann:2013} 
(their Fig.~16) shows the improvements achieved with the new pipeline.

\subsection{Limiting sensitivity and signal-to-noise}\label{sect:depth}

\begin{figure}
 \resizebox{\hsize}{!}{\includegraphics{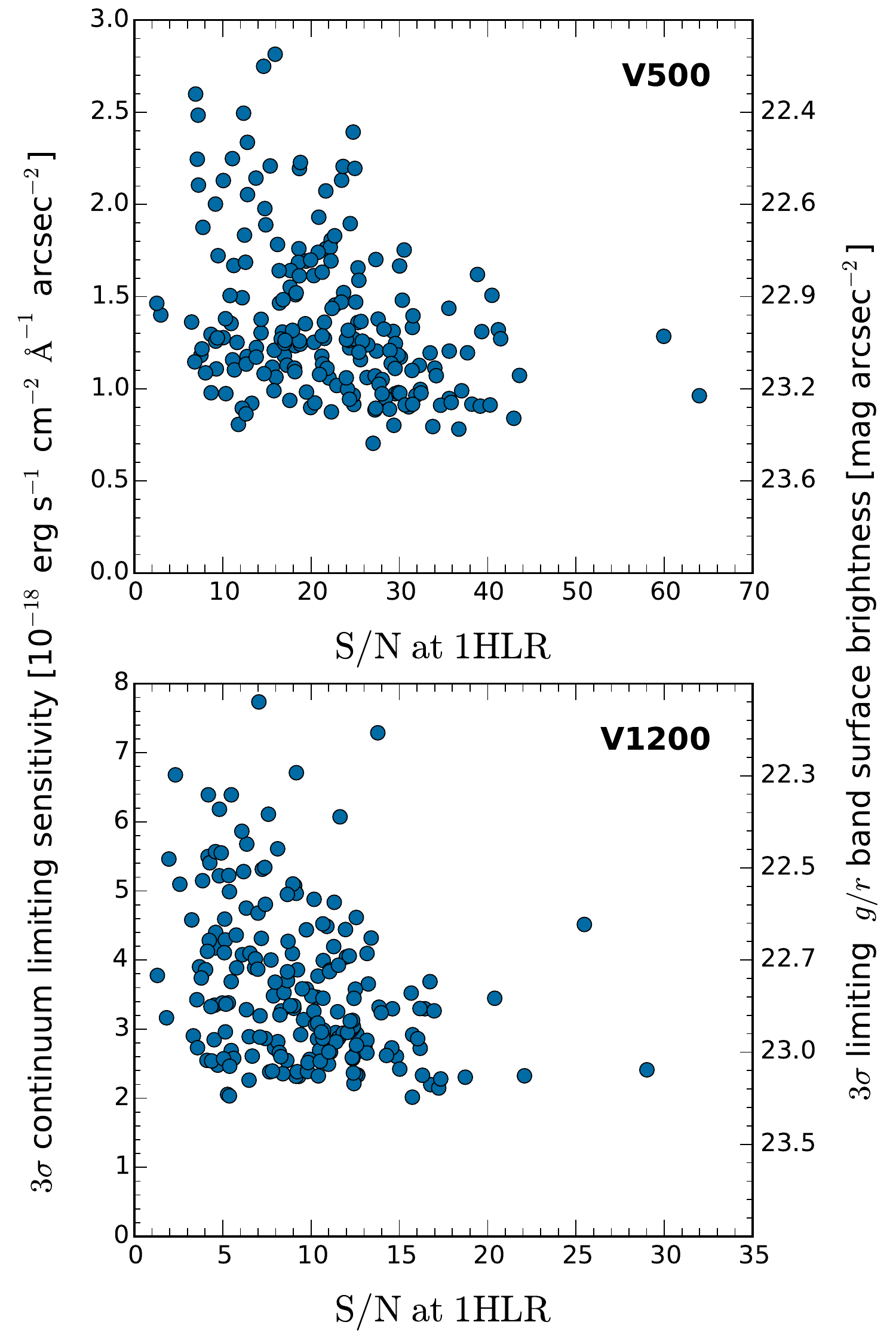}}
 \caption{Limiting $3\sigma$ continuum sensitivity as a function of the average continuum S/N 
	at the half light radius (HLR). The corresponding broad-band surface brightness limits 
   	in $r$ (V500) and $g$ (V1200) are indicated on the right $y$-axis. The limiting continuum 
	sensitivity and the S/N were computed from the median signal and noise in the wavelength 
	region 4480--4520\AA\ and 5590--5680\AA\ for the V1200 and V500 data, respectively.}
 \label{fig:DR2_depth}
\end{figure}

In order to assess the depth of the data, we estimated the 3$\sigma$ continuum flux density detection 
limit per interpolated $1\,\mathrm{arcsec}^{2}$-spaxel for the faintest regions. Fig.~\ref{fig:DR2_depth} 
shows the limiting continuum sensitivity of the spectrophotometrically recalibrated CALIFA spectra. The 
depth is plotted against the average S/N per spectral resolution element within an elliptical 
annulus of $\pm1\arcsec$ around the galaxies $r$-band half-light semi-major axis (HLR), with PA 
and radius values taken from \citetalias{Walcher:2014}. 
A narrow wavelength window at 4480--4520\AA\ for the V1200 and at  5590--5680\AA\ for the V500 
was used to estimate both values. These small windows are nearly free of stellar absorption features 
or emission lines. The 3$\sigma$ continuum flux density detection 
limit\footnote{Note that this is a continuum flux density. See Note 5 of \citetalias{Husemann:2013}.} 
for the V1200 data 
($I_{3\sigma}=3.2\times10^{-18}\,\mathrm{erg}\,\mathrm{s}^{-1}\,\mathrm{cm}^{-2}\,\mathrm{\AA}^{-1}\,\mathrm{arcsec}^{-2}$
in the median at 4500\AA) is a factor of $\sim$2-3 brighter than for the V500 data
($I_{3\sigma}=1.2\times10^{-18}\,\mathrm{erg}\,\mathrm{s}^{-1}\,\mathrm{cm}^{-2}\,\mathrm{\AA}^{-1}\,\mathrm{arcsec}^{-2}$
in the median at 5635\AA) 
mainly due to the difference in spectral resolution. 
These continuum sensitivities can be transformed into equivalent limiting broad band surface brightnesses 
of $23.0\,\mathrm{mag}\,\mathrm{arcsec}^{-2}$ in the $g$-band for the V1200 data and
$23.4\,\mathrm{mag}\,\mathrm{arcsec}^{-2}$ in the $r$-band for the V500.
The variance of the sky brightness of each night might be one of the main factors 
of the dispersion in the limiting continuum sensitivity.
Dust attenuation, transparency of the night, and other atmospheric conditions might also affect 
the achievable depth at fixed exposure times. 

The limiting sensitivity is a measure of the noise and thus it correlates mildly with the S/N.
The mean S/N in the continuum per spaxel at the half-light semi-major axis (HLR) of all objects is 
$\sim$9.5 for the V1200 setup, while it is $\sim$22.2 for the V500 data. Thus, we achieve a 
S/N$\gtrapprox$10 for a significant number of the objects even for the V1200 setup. 


\section{Access to the CALIFA DR2 data}\label{sect:DR2_access}

\subsection{The CALIFA DR2 search and retrieval tool}

The public data is distributed through the CALIFA DR2 web page (\url{http://califa.caha.es/DR2}). A simple web 
form interface, already in use for the first data release, allows the user to select data of a particular 
target galaxy, or a subsample of objects within some constraints on observing conditions or galaxy properties. 
Among the selection parameters we include the instrument setup, galaxy coordinates, redshift, $g$-band 
magnitudes, observing date, Hubble type, bar strength, inclination estimated from axis ratio, V band 
atmospheric attenuation, airmass, and relative accuracy of the SDSS/CALIFA photometric calibration. 

If any CALIFA datasets are available given the search parameters, they are listed in the follwing web 
page and can be selected to be downloaded. The download process requests a target directory on the local 
machine to store the data, after the downloading option was selected. The CALIFA data are delivered as 
fully reduced datacubes in FITS format separately for each of the two CALIFA spectral settings, i.e. the 
V500 and V1200 setup. Each DR2 dataset is uniquely identified by their file name, 
\texttt{$GALNAME$.V1200.rscube.fits.gz} and \texttt{$GALNAME$.V500.rscube.fits.gz} for the V1200 and V500 
setup respectively, where $GALNAME$ is the name of the CALIFA galaxy listed in Table \ref{tab:DR2_sample}.

All the QC tables discussed along this article are also distributed in CSV and FITs-table formats in the
same webpage. In addition, we distribute the more relevant tables discussed in \citetalias{Walcher:2014} 
regarding the characterization of the MS, using similar formats. These tables could be useful in further
science explorations of the cubes.

\subsection{Virtual Observatory services}

CALIFA data is also available through several Virtual Observatory (VO) facilities.

\begin{enumerate}
\item The FITS files of the full cubes are accessible through GAVO's ObsCore \citep{Louys:2011} 
service, which is part of the TAP \citep{Dowler:2011} service at \url{http://dc.g-vo.org/tap}. 
ObsCore provides a homogeneous description of observational data products of all kinds and thus 
allows global data set discovery.  The system already supports the upcoming IVOA DataLink 
standard for performing cutouts and similar server-side operations.

\item At the same TAP endpoint, the \texttt{califadr2.cubes} and \texttt{califadr2.objects} 
tables allow queries versus CALIFA-specific metadata, in particular the quality control parameters 
given in Tables~\ref{tab:QC_par_V500} and \ref{tab:QC_par_V1200}.

\item Individual, cut-out spectra can be located and retrieved from the CALIFA SSA 
service\footnote{SSA access URL \url{http://victor:8080/califa/q2/s/info}.}; advanced SSAP 
clients like Splat \citep{Draper:2014} also support server-side spectral cutouts 
on this service via a DataLink prototype.

\item The spaxels can also be queried in database tables via GAVO's TAP service mentioned 
above (the tables are called \texttt{califadr2.fluxv500} and \texttt{califadr2.fluxv1200}).
\end{enumerate}

An overview of VO-accessible resources generated from CALIFA -- possibly updated from what is 
reported here -- is available at \url{http://dc.g-vo.org/browse/califa/q2}. This page also gives 
some usage scenarios for CALIFA VO resources.


\section{Conclusions}

Along this article we have presented the main characteristics of the second public data release
of the Calar Alto Legacy Integral Field Area (CALIFA) survey. This data release comprises 
200 galaxies (400 datacubes) containing more than 1.5 million spectra\footnote{Obtained from 
$\sim$ 400000 independent spectra from the RSS files.}, covering a wide range of masses, 
morphological types, colors, etc. This subset of randomly selected objects conforms a 
statistically representative sample of the galaxies in the Local Universe.
The CALIFA DR2 provides science-grade and quality-checked integral-field spectroscopy publicly 
distributed to the community at \url{http://califa.caha.es/DR2}.

We have described in detail the main quality parameters analysed in the validation process, provided
to the users with complete tables to select the objects for their science cases. The data have 
been reduced using a new version of the pipeline (V1.5), which improves considerable the quality
of the data in terms of: (i) the spatial resolution, (ii) the covariance between the adjacent
spectra, and (iii) the spectrophotometric calibration.

Compared with other on-going major surveys, CALIFA offers a better spatial resolution. 
The PSF of the datacubes has been improved considerably, with a mean value of $\sim$ 2.5\arcsec\ 
(Sect.\ref{sect:spatial_resolution}), similar to SAMI \citep{Sharp:2014}. In the case of MaNGA, 
the combination of an average seeing at the Sloan Telescope ($\sim$ 1.5\arcsec) and the fiber 
size (2\arcsec), would produce a PSF with a very similar FWHM. The redshift range of SAMI and 
MaNGA surveys is considerably larger than of CALIFA, reaching up to $z\sim$0.1. 
This means that only for galaxies at the lowest redshift range SAMI and MaNGA will offer a 
similar physical resolution. On the other hand, the spatial coverage of CALIFA is larger than any
of those surveys, both in physical and in projected terms (five times larger than SAMI and two times 
larger than MaNGA). In summary, CALIFA is the survey that samples the galaxies with the largest number
of spatial elements for the largest FoV. The penalty for this wider coverage is
the lower number of galaxies observed (6 times lower than SAMI and 15 times lower than MaNGA), 
and a lower spectral resolution of CALIFA in the full wavelength range.

The shorter redshift range, covering 300 Myrs in cosmological times, provides homogeneity of 
the derived properties, such as the SFH or the gas abundances, which in other surveys should be 
corrected prior to making an homogeneous comparison of their full sample.

The dataset analysed so far have produced significant advances in our knowledge of the stellar 
and gas composition in galaxies, their kinematical structure, and the overall star formation history
and chemical enrichment (as reviewed in  the introduction). We have uncovered new local relations 
within galaxies, tightly connected to the global ones described using classical spectroscopic 
surveys. With this new DR we open to the astronomical communitty the posibility to futher analyse 
the spatially resolved properties of galaxies, presenting a panoramic view of the galaxy properties.

\begin{acknowledgements}
CALIFA is the first legacy survey being performed at Calar Alto. The CALIFA collaboration would like to thank 
the IAA-CSIC and MPIA-MPG as major partners of the observatory, and CAHA itself, for the unique access
to telescope time and support in manpower and infrastructures. The CALIFA collaboration thanks also the CAHA 
staff for the dedication to this project. 
RGB, RGD and EP are supported by the Spanish {\it Ministerio de   Ciencia e Innovaci{\'o}n} under grant AYA2010-15081.
SZ has been supported by the EU Marie Curie Integration Grant ``SteMaGE'' Nr. PCIG12-GA-2012-326466 
(Call Identifier: FP7-PEOPLE-2012 CIG).
JFB acknowledges support from grants AYA2010-21322-C03-02 and AIB-2010-DE-00227 from the Spanish Ministry of 
Economy and Competitiveness (MINECO), as well as from the FP7 Marie Curie Actions of the European Commission, 
via the Initial Training Network DAGAL under REA grant agreement number 289313.
Support for L.G. is provided by the Ministry of Economy, Development, and Tourism's Millennium Science Initiative 
through grant IC12009, awarded to The Millennium Institute of Astrophysics, MAS. L.G. also acknowledges support 
by CONICYT through FONDECYT grant 3140566.
AG acknowledges support from the FP7/2007-2013 under grant agreement n. 267251 (AstroFIt).
JMG acknowledges support from the Funda\c{c}\~{a}o para a Ci\^{e}ncia e a Tecnologia (FCT) through the 
Fellowship SFRH/BPD/66958/2009 from FCT (Portugal) and research grant PTDC/FIS-AST/3214/2012.
RAM was funded by the Spanish programme of International Campus of Excellence Moncloa (CEI).
JMA acknowledges support from the European Research Council Starting Grant (SEDmorph; P.I. V. Wild).
IM, JM and AdO acknowledge the support by the projects AYA2010-15196 from the Spanish Ministerio de Ciencia e 
Innovaci\'on and TIC 114 and PO08-TIC-3531 from Junta de Andaluc\'ia.
AMI acknowledges support from Agence Nationale de la Recherche through the STILISM project (ANR-12-BS05-0016-02).
MM acknowledges financial support from AYA2010-21887-C04-02 from the Ministerio de Economía y Competitividad.
PP is supported by an FCT Investigador 2013 Contract, funded by FCT/MCTES (Portugal) and POPH/FSE (EC). He acknowledges 
support by FCT under project FCOMP-01-0124-FEDER-029170 (Reference FCT PTDC/FIS-AST/3214/2012), funded by FCT-MEC (PIDDAC) 
and FEDER (COMPETE). 
TRL thanks the support of the Spanish Ministerio de Educaci\'on, Cultura y Deporte by means of the FPU fellowship.
PSB acknowledges support from the Ram\'on y Cajal program, grant ATA2010-21322-C03-02 from the Spanish 
Ministry of Economy and Competitiveness (MINECO).
CJW acknowledges support through the Marie Curie Career Integration Grant 303912.
V.W. acknowledges support from the European Research Council Starting Grant (SEDMorph P.I. V.~Wild) 
and European Career Re-integration Grant (Phiz-Ev P.I. V.~Wild). 
YA acknowledges financial support from the \emph{Ram\'{o}n y Cajal} programme (RyC-2011-09461) and project 
AYA2013-47742-C4-3-P, both managed by the \emph{Ministerio de Econom\'{i}a y Competitividad}, as well as the 
`Study of Emission-Line Galaxies with Integral-Field Spectroscopy' (SELGIFS) programme, funded by the 
EU (FP7-PEOPLE-2013-IRSES-612701) within the Marie-Sklodowska-Curie Actions scheme. 

\end{acknowledgements}

\bibliographystyle{aa}
\bibliography{CALIFA_DR2}

\appendix
\section{Computing the error spectrum for co-added spectra}\label{ap:correlation}

Some science cases require a minimum S/N in the spectra, especially in the outer parts of the 
galaxies. This is achived by spatially co-adding spaxels in the datacubes, often by means 
of an adaptive binning method like the Voronoi-binning scheme, implemented for optical 
IFS data by \citet{Cappellari:2003}. However, the final error spectrum of the co-added spectra 
cannot be simply quadratically summed since the spectra are not independent of each other. 
As described in Sect.~\ref{sect:pipeline}, we adopt an inverse-distance weighted image 
reconstruction which, like many other image resampling schemes, introduces a correlation 
between spaxels in the final datacube. In Sect.~\ref{sect:correlation} we provide 
an equation that relates the analytically propagated error recorded in the datacubes with the 
final ''real`` error of the co-added spectra\footnote{See also Sect. 3.2 and 3.3 of 
\citet{CidFernandes:2013} for a detailed disquisition on error propagation and correlated 
noise for IFS.}.

Let $B$ be a bin of size $N$ spectra, i.e. we want to co-add $N$ spectra and 
compute the corresponding error spectrum for that bin. Since we are adding the flux 
to obtain an integrated spectra, first we need to add the errors of 
each individual spectra in quadrature:

\begin{equation*}
\label{eq:error_lambda}
\epsilon_{B}^2 = \sum\limits_{k=1}^{N} \epsilon_{k}^2 
\end{equation*}

This would be the error spectrum of the bin $B$ if the spaxels where completely independent. 
To account for the correlated noise, we just need to multiply by the corresponding 
``correlation factor'' (Eq.~\ref{eq:correlation}) for a given number of spectra in a 
particular bin:

\begin{equation*}
 \epsilon^{2}_\mathrm{real,B} = \beta (N)^{2} \times \epsilon_{B}^2
\end{equation*}

When the bin $B$ contains a large number of spaxels ($N$ $\gtrapprox$ 80), the use of 
Eq.~\ref{eq:correlation} is not recommended. 
In this case, the ERRWEIGHT HDU extension of the CALIFA FITS file datacube should be 
used (see Table \ref{tab:HDUs}) as a correction factor for each spaxel.

\end{document}